\documentclass[twocolumn,floatfix,nofootinbib]{revtex4-2}
\usepackage[utf8]{inputenc}
\usepackage[T1]{fontenc}

\usepackage[labelfont=bf, textfont=bf]{caption}
\usepackage{subcaption}
\usepackage[usenames]{color}
\usepackage{float}
\usepackage{bm}

\usepackage[english]{babel}
\usepackage{url}
\usepackage{amsmath}
\usepackage{graphicx}
\usepackage{parskip}
\usepackage{fancyhdr}
\usepackage{booktabs}
\usepackage[bottom]{footmisc}
\usepackage{enumerate}
\usepackage{bold-extra}
\usepackage{textgreek}
\usepackage{verbatim}
\usepackage{placeins}
\usepackage{soul}
\usepackage{adjustbox}
\usepackage{multirow}
\usepackage{diagbox}
\usepackage{refcount}
\usepackage{hhline}
\usepackage{amssymb}
\usepackage{pifont}

\providecommand{\eb}[1]{\textcolor{black}{#1}}

\providecommand{\brkt}[1]{\ensuremath{\left[ #1 \right]}}
\providecommand{\angl}[1]{\ensuremath{\left\langle #1 \right\rangle}}

\providecommand{\dertot}[2]{\frac{\textmd{d} #1}{\textmd{d} #2}}
\providecommand{\dertotinline}[2]{\textmd{d} #1/ \textmd{d} #2}

\providecommand{\derpar}[2]{\frac{\partial #1}{\partial #2}}

\begin{document}
	
\title{Effect of injection conditions on the non-linear behavior of the ECDI and related turbulent transport}
\author{E. Bello-Benítez, A. Marín-Cebrián and E. Ahedo }
\affiliation{Department of Aerospace Engineering, Universidad Carlos III de Madrid, Leganés, Spain}
\date{\today}

\begin{abstract}
    The electron-cyclotron drift instability (ECDI) has been proposed as one of the main actors behind the anomalous transport of electrons in Hall plasmas. 
    In this work, we revisit the classical theory of this instability [Forslund et al., Phys. Rev. Lett. 25, 1266 (1970)] and perform two-dimensional kinetic simulations under several conditions to analyze the non-linear behavior and the induced transport.
    Fully-periodic simulations, with conditions faithful to the linear theory are analyzed first. In agreement with existing literature, they show the growth of ECDI modes, ion-wave trapping vortexes and an induced cross-field electron current in early simulation times.
    However, in contrast with similar works, non-linear saturation is observed and the plasma tends, in the long term, to a new equilibrium with mild oscillations and mild anomalous current. 
    This evolution is consistent to what can be expected from energy conservation.
    The quenching of the oscillations seem to be highly related with the distortion of ion vortexes in phase space after a long-term interaction with the electrostatic wave. 
    This result suggests that sustained oscillations and turbulent current could thrive if ions are renewed by, e.g., removing and injecting particles through axial boundaries instead of applying periodicity.
    This second type of simulations shows that injection conditions highly impact the late simulation behavior of ECDI oscillations, where we identify several regimes depending on the value of the ion residence time compared to the characteristic saturation time in the fully periodic case.
    The intermediate regime, where these two times are close, is the only one providing sustained oscillations and electron transport.
\end{abstract}
\maketitle


\section{Introduction}

The problem of anomalous electron cross-field transport remains as one of big open challenges for the community of $\bm E \times \bm B$ plasmas. 
In the field of plasma propulsion, this problem has been mainly studied in the context of Hall-thruster discharges and represent one big obstacle on the way towards predictive efficient numerical models.
The large drift of electrons in the azimuthal direction of the Hall thruster is a source of several families of azimuthal oscillations that are potential candidates to explain the anomalous transport and have been observed experimentally \cite{chou01,elli12,macd11,tsik09b,tsik15,tsik19c}. 
The classical explanation \cite{jane66} for the impact of oscillations on transport relies on the correlation of oscillations in density and electric field in the $\bm E \times \bm B$ direction under the presence of a magnetic field.


When using a kinetic formulation for the electrons, the analytical studies   of instabilities are usually limited to a homogeneous and collisionless plasma.
For the conditions of a Hall plasma, where electrons are magnetized but ions are not, the dispersion relation of the classical electron cyclotron drift instability (ECDI) is obtained \cite{fors70,wong70}. 
This classical instability has been revisited, during the last two decades, by several authors \cite{adam04,ducr06,cava13,lafl16b,janh18b}. 

Kinetic models aimed to analyze electron turbulent transport in Hall plasmas in the plane perpendicular to the applied magnetic field $\bm B_0$ can be classified in 1D-azimuthal \cite{janh18,lafl16a,tacc19c} and 2D-axial-azimuthal \cite{adam04,lafl17b,coche14}. 
Many 2D simulations include a number of phenomena that makes challenging to compare the results with the ECDI linear theory; such as, inhomogeneous magnetic field, collisions, ionization or electrical connection between anode and cathode.
The 1D-azimuthal simulations are closer to the linear theory of the ECDI but some of them still add effects that are not considered in the dispersion relation, such as refreshing of particle velocities or collisions. 
The 1D simulation results reported in the literature have been seen to be very sensitive to the axial treatment of particles; observing a sustained oscillatory saturated state only when using a virtual axial dimension and particle velocities are refreshed after leaving the domain axially \cite{lafl16a, tacc19c, smol19c}. 
If particles are not refreshed an unbounded growth of oscillations and heating are seen \cite{lafl16a,janh18,smol23}.

\eb{In this work we present fundamental simulation results of scenarios close to the theoretical conditions leading to the ECDI, but not directly related to the far more complex Hall-thruster discharge.
However, the understanding of the simplified scenarios can give a valuable physical insight that is key to analyze more realistic scenarios where many other effects can interplay.
}
In section \ref{sec: dispersion relation}, we revisit the main results from the classical dispersion relation of the ECDI theory.
A recently developed in-house two-dimensional particle-in-cell (PIC) code, called PICASO (Particle-In-Cell Advanced SimulatOr), is introduced in section \ref{sec: periodic} and used to analyze the non-linear evolution of the ECDI under several conditions.
In every case, a 2D domain perpendicular to $\bm B_0$ is considered; which we will refer to as the axial-azimuthal plane, making the analogy with a Hall-discharge geometry.
In this section, we consider a fully periodic domain and simulate under the assumptions of the classical ECDI theory.
\eb{These results can be compared with the existing literature \cite{janh18,smol23,lafl16a} on 1D azimuthal simulations of the classical ECDI without refreshing, but open the possibility to the growth of modes with an axial component. 
Our results are in good agreement with those shown elsewhere during early simulation times; but in the long-term non-linear saturation is reached even without particle refreshing, which is not shown in the previous literature.
A stationary state is reached that shows no oscillations capable of driving electron transport, which is the only possible stationary behavior according to the energy balance.
The periodic simulation results suggest that the quenching of oscillations is related to the distortion of ion vortexes in phase space.}

Periodic results point to a relation between ion behavior and the existence of electron transport in the long term, what suggests that renewal of ion particles could help to develop sustained ECDI oscillations and axial transport.
This motivate us to replace axial boundary conditions from periodic to removal/injection of particles in section \ref{sec: injection}. 
\eb{
Our motivation is, thus, different from that of previous works on 1D-azimuthal simulations that use axial re-injection (i.e., conserving particles) and velocity refreshing \cite{lafl16a,tacc19c} to reach saturation and to limit heating; which we observe even with full-periodic conditions.
Our approach is also different; we decided to inject particles with fixed fluxes to mimic the generation of particles in a finite plasma such as the Hall discharge.
Moreover, since the axial direction is not included in previous 1D-azimuthal models, the axial transport of particles and waves is not completely resolved and their refreshing strategy is equivalent to having inelastic collisions \cite{tacc19c}.
Our model is 2D, including the axial direction, so that injection cannot longer be reduced to a collisional phenomena and the mixing of injected and bulk populations is self-consistently resolved.
}

\eb{As one of the key conclusions of this work, we see how the axial transport of ions determines the long-term behavior of the oscillations and
we are able to establish a clear relation between the ion velocity (related with the residence time) and the long-term behavior of sustained electron transport induced by short-wavelength oscillations.
}

\section{The classical ECDI: linear analysis}
\label{sec: dispersion relation}

We attempt to study the stability of an homogeneous, collisionless  plasma at equilibrium subjected to mutually perpendicular magnetic
$\bm B_0 = B_0 \bm 1_x$ and electric $\bm E_0 = E_0 \bm 1_z$  fields.
Throughout the article subindex `0' and `1' stand for equilibrium and perturbed conditions.
Consistent with Hall thruster discharges: the strength of the stationary magnetic field  keeps electrons well-magnetized and ions unmagnetized, and the plasma currents are low enough to neglect the contribution of  
the self-induced magnetic field $\bm B_1$.
At equilibrium, electrons have a drifted-Maxwellian velocity distribution function (VDF) with density $n_0$, temperature $T_{e0}$, and an $\bm E_0 \times \bm B_0$ drift velocity $\bm u_{e0} = u_{ye0} \bm 1_y$, $u_{ye0} = E_0/B_0$. 
Equilibrium ions are assumed cold, have a density $n_0$, and move with a constant velocity $\bm u_{i0} = u_{zi0} \bm 1_z$, parallel to $\bm E_0$. 
This homogeneous equilibrium disregards the electrostatic acceleration of ions due to $\bm E_0$. 
From analogy with a Hall-thruster geometry, let us refer throughout the article to the directions of $\bm B_0$, $\bm E_0$ and $\bm E_0 \times \bm B_0$ as radial ($x$), axial ($z$) and azimuthal ($y$).

Next, we summarize the main aspects of the well-known linear stability analyses \cite{fors70,wong70,ducr06,cava13,lafl16b}.
From the linearly perturbed Vlasov equation, the relation between the electron density and the electric potential perturbations 
of wavevector $\bm k$ and complex frequency $\omega$, for propagation perpendicular to $\bm B_0$  (i.e., the component of  $\bm k$ parallel to $\bm B_0$ is zero),
is
\begin{equation}
\frac{n_{e1}}{n_0} =\Big[1-g(\omega_e,b_e;\omega_{ce})\Big]\frac{e \phi_1}{T_{e0}}
\label{eq: perturbation electrons}
\end{equation}
with
\begin{equation}
    g(\omega_e,b_e;\omega_{ce})=\exp(-b_e)\Big[I_0(b_e)+ 2 \sum_{m=1}^{\infty} \frac{\omega_e^2  I_m(b_e)}{ \omega_e^2 - m^2 \omega_{ce}^2}\Big]. 
\label{eq: g}
\end{equation}
In these equations 
$\bm E=-\nabla \phi$,
$c_{e0} = \sqrt{T_{e0}/m_e}$ is the electron thermal velocity, 
$\omega_{ce} = eB_0/m_e$ is the cyclotron frequency,  
$b_e = k^2 \rho_{e0}^2$,
$\rho_{e0} = c_{e0}/\omega_{ce}$ is the electron Larmor radius,  
$\omega_e = \omega - k_y u_{ye0}$ is the electron Doppler-shifted frequency, 
$\bm k = k_y \bm 1_y + k_z \bm 1_z$,
and
$I_m$ are the modified Bessel functions of the first kind.
According to equation \eqref{eq: g}, the perturbed electron response has resonances at the electron gyrofrequency and its harmonics, i.e. $\omega_e=m\omega_{ce}$.

Next, the density perturbations of cold ions follow 
\begin{equation}
    \frac{n_{i 1}}{n_0}=\frac{k^2 c_{s0}^2}{\omega_{i}^{2}} \frac{e \phi_1}{T_{e0}},
    \label{eq: perturbation ions}
\end{equation}
with $\omega_i = \omega - k_z u_{zi0}$ the ion Doppler-shifted frequency, and  $c_{s0} = \sqrt{T_{e0}/m_i}$ the ion sound speed.
Using Eqs. \eqref{eq: perturbation electrons} and \eqref{eq: perturbation ions}, together with the linearized Poisson equation, yields the two-dimensional dispersion relation
\begin{equation}
       1 + k^{2} \lambda_{D0}^{2} = \frac{k^{2} c_{s0}^{2}}{\omega_{i}^{2}} + g(\omega_e,b_e;\omega_{ce}).
           \label{eq: ECDI}
\end{equation}
This equation is solved for the complex frequencies $\omega = \omega_r + \text{i} \gamma$ with all other parameters fixed, including $\bm k$. 

\begin{table*}[!t]
\centering
{\renewcommand{\arraystretch}{1.4}
\begin{tabular}{|c|c|c|}
\hline
\textbf{Type}                                                                                           & \textbf{Description and symbol}                  & \textbf{Value and units} \\ \hline \hline
\multirow{8}{*}{\textbf{\begin{tabular}[c]{@{}c@{}}Fundamental \\ plasma parameters\end{tabular}}}      & Ion mass, $m_i$                                  & 1 u                      \\ \cline{2-3} 
                                                                                                        & Electric field, $E_0$                            & $10^4$ V/m               \\ \cline{2-3} 
                                                                                                        & Magnetic field, $B_0$                            & 200 G                    \\ \cline{2-3} 
                                                                                                        & Plasma density, $n_0$                            & $10^{17}$ m$^{-3}$       \\ \cline{2-3} 
                                                                                                        & Ion axial velocity, $u_{zi0}$                    & 2.5 km/s                 \\ \cline{2-3} 
                                                                                                        & Electron temperature, $T_{e0}$                   & 6 eV                     \\ \cline{2-3}  \hline \hline
\multirow{9}{*}{\textbf{\begin{tabular}[c]{@{}c@{}}Derived \\ plasma parameters\end{tabular}}}          & Electron azimuthal drift, $u_{ye0}$              & 500 km/s                 \\ \cline{2-3} 
                                                                                                        & Electron thermal speed, $c_{e0}$                 & 1027 km/s                \\ \cline{2-3} 
                                                                                                        & Sound speed, $c_{s0}$                              & 23.97 km/s               \\ \cline{2-3} 
                                                                                                        & Debye length, $\lambda_{D0}$                     & 57.58 $\mu$m             \\ \cline{2-3} 
                                                                                                        & Electron azimuthal-drift gyroradius, $\ell_{e0}$ & 142.1 $\mu$m             \\ \cline{2-3} 
                                                                                                        & Electron Larmor radius, $\rho_{e0}$              & 292.0 $\mu$m             \\ \cline{2-3} 
                                                                                                        & Electron plasma frequency, $\omega_{pe0}$        & 2.839 GHz                \\ \cline{2-3} 
                                                                                                        & Electron gyrofrequency, $\omega_{ce}$            & 0.5600 GHz               \\ \cline{2-3} 
                                                                                                        & Ion plasma frequency, $\omega_{pi0}$             & 66.26 MHz                \\ \cline{2-3} 
                                                                                                        & Lower-hybrid frequency, $\omega_{lh}$            & 13.07 MHz                \\ \hline \hline
\multirow{6}{*}{\textbf{\begin{tabular}[c]{@{}c@{}}Fundamental \\ numerical parameters\end{tabular}}}   & Azimuthal domain length, $L_y$                   & 5.359 mm                 \\ \cline{2-3} 
                                                                                                        & Axial domain length, $L_z$                       & 2.679 mm                 \\ \cline{2-3} 
                                                                                                        & Number of azimuthal celss, $N_y$                 & 100                      \\ \cline{2-3} 
                                                                                                        & Number of axial cells, $N_z$                     & 50                       \\ \cline{2-3} 
                                                                                                        & Number of particles per cell, $N_{\textmd{ppc}}$ & 200                      \\ \cline{2-3} 
                                                                                                        & Time steps, $\Delta t$                           & $5 \times 10^{-12}$ s    \\ \cline{2-3} 
                                                                                                        & Number of time steps, $N_t$                      & $6 \times 10^5$          \\ \cline{2-3} 
                                                                                                        & Number of time steps between print-outs, $N_{\textmd{print}}$      & 1000             \\ \hline \hline
\multirow{2}{*}{\textbf{\begin{tabular}[c]{@{}c@{}}Derived \\ numerical parameters\end{tabular}}}     & Azimuthal cell size, $\Delta y$                  & 53.59 $\mu$m             \\ \cline{2-3} 
                                                                                                        & Axial cell size, $\Delta z$                      & 53.59 $\mu$m             \\ \hline
\end{tabular}
}
\caption{Physical and numerical parameters of the reference simulation case. The subscript `0' stands for initial equilibrium conditions. Derived parameter values are included for completeness but can be computed from fundamental ones.}
\label{tab: parameters}
\end{table*}

The solutions of the dispersion relation are pairs of modes of three types \cite{wong70}.
First, there is a pair of ion-acoustic modes with $g \ll k^2 c_{s0}^2/\omega_i$ and real-valued frequencies 
$\omega_i = \pm \omega_{\textmd{IA}}$; being
    \begin{equation}
         \omega_{\textmd{IA}} = \frac{k c_{s0}}{ \sqrt{1 + k^2 \lambda_{D0}^2 }},
         \label{eq: ion acoustic}
    \end{equation}
which include a non-neutral term in the denominator that corrects the quasineutral linear relation $\omega_i = \pm k c_{s0}$.
Second, there are  electron Bernstein waves \cite{swan03},  with $g \gg k^2 c_{s0}^2/\omega_i$ and real-valued frequencies too, close to the resonances $\omega_e=m\omega_{ce}$.
Third, when $g \sim k^2 c_{s0}^2/\omega_i$,  Bernstein waves become coupled with the ion acoustic modes,
yielding one of them the modified ion acoustic (MIA) pair
with frequencies 
    \begin{equation}
        \omega_i = \pm \frac{k c_{s0}}{ \sqrt{1 + k^2 \lambda_{D0}^2 - g(\omega_e)}}. 
        \label{eq: omega_i}
    \end{equation}
    
For $g>1 + k^2 \lambda_{D0}^2$, one of MIA modes is unstable leading to the so called ECDI.
Since $g$ changes from $-\infty$ to $+\infty$ when crossing the resonance $\omega_e=m\omega_{ce}$ with $\omega_e$ increasing, the MIA mode starts becoming unstable at 
$\omega_e=(m\omega_{ce})^+$ and becomes stable before 
reaching the $(m+1)$ resonance \cite{ducr06, cava13, lafl16b}. 

\begin{figure}[t]
    \centering
    \begin{subfigure}[b]{0.49\textwidth}
        \caption{}
        \centering
        \includegraphics[width=\textwidth]{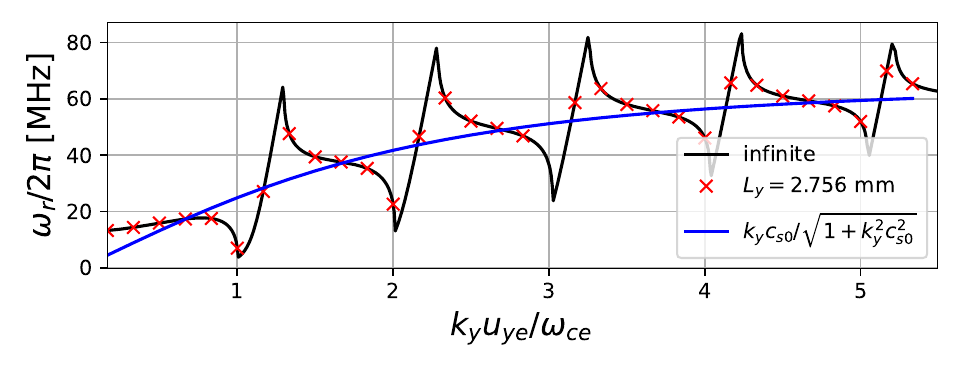}
    \end{subfigure}
    \begin{subfigure}[b]{0.49\textwidth}
        \caption{}
        \centering
        \includegraphics[width=\textwidth]{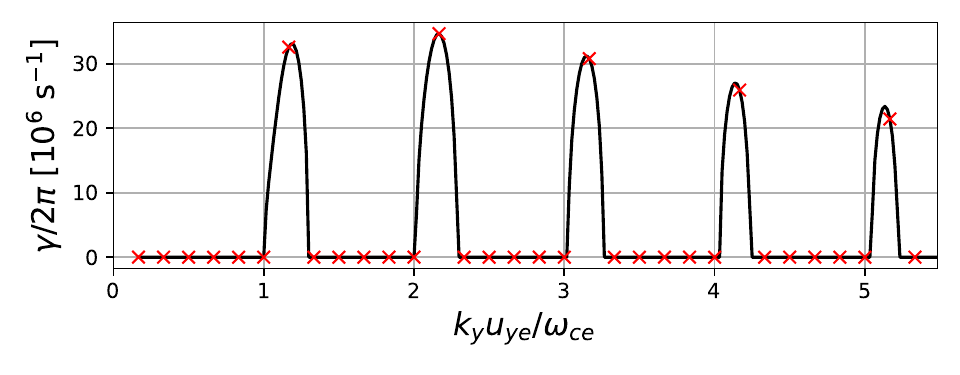}
    \end{subfigure}
    \caption{ECDI dispersion relation for a hydrogen plasma  with $k_\parallel = 0$, $k_z = 0$ and the equilibrium conditions summarized in table \ref{tab: parameters}. The black solid line is the solution for an infinite plasma. Red crosses stand for the discrete solutions in a finite plasma with $L_y = 5.359$ mm.
    The blue solid line is the ion-acoustic frequency  $\omega_{\textmd{IA}}$.
    The dispersion relation has been solved with the same numerical approach than reference \cite{cava13}.
    }
    \label{fig: dispersion relation}
\end{figure}

Figure  \ref{fig: dispersion relation} depicts, for the  equilibrium solution of  Table \ref{tab: parameters}, the complex frequency of the purely-azimuthal (i.e. $\bm k=k_y \bm 1_y$) MIA mode, and shows the growth rate of the ECDI linked to each $m\omega_{ce}$;
as a function of $k_y \ell_{e0}$, being $\ell_{e0} = u_{ye0}/\omega_{ce}$ the azimuthal-drift gyroradius. 
The ion-acoustic frequency $\omega_{\textmd{IA}}$ is also plotted for comparison with the MIA mode.
The same equilibrium solution will be used in the kinetic simulations of the coming sections.
For $k_y > 0$, the unstable mode has $\omega_r > 0$, meaning propagation in the $\bm E_0 \times \bm B_0$ direction.
Due to symmetry, for $k_y < 0$, the unstable mode  has $\omega_r < 0$ (i.e.,  moving again along $\bm E_0 \times \bm B_0$) and same $\gamma$. 
In the purely azimuthal ECDI, one has $\omega\sim k_yc_s \ll \omega_e\simeq -k_yu_{ye0}\approx -m\omega_{ce}$. Therefore, the electron (Bernstein-type) response is quasi-steady and defines mostly the wavelength of the ECDI.
Regarding nonneutral effects the MIA is quasineutral as long as $k_y^2\lambda_{D0}^2\ll 1$, i.e.
\begin{equation}
    \dfrac{\lambda_{D0}}{\rho_e}\ll \dfrac{u_{y_{e0}}}{m c_{e0}}
\end{equation}
and non-neutral effects tend to reduce the complex frequency.
This effect is very clearly seen in the curve  $\omega_{\textmd{IA}}$ in figure \ref{fig: dispersion relation}(a), that shows frequencies lowered with respect to a  quasineutral ion-acoustic linear relation. 
The comparison of the MIA and ion-acoustic frequencies demonstrates that they follow the same trend but with significant deviations coming from the coupling with the Bernstein terms.

If the plasma has a finite size $L_y$ along $y$, 
there is only a 
discrete wave spectrum with  $k_y = n 2 \pi /L_y$ and $n$  the number of wavelengths fitting in the domain.
Red crosses in Fig.  \ref{fig: dispersion relation} show that spectrum for $L_y =12\pi \ell_{e0}= 5.359$ mm,
when resonances correspond to $n$ a multiple of 6.
That length has been chosen so that modes $n=1+6 m$ capture approximately the peaks in $\gamma$ associated to each resonance $m$.
For the chosen parameters, the fastest growing mode is $n = 13$, close to the $m=2$ resonance, with {$\gamma/2\pi = 34.7 \times 10^6$ s$^{-1}$},  $k_y \ell_{e0} = 2.167$ and  $\omega_r = 46.6$ MHz. 
In terms of growth rate, the mode $n = 7 $, in the band $m=1$, follows closely with {$\gamma/2\pi = 32.6 \times 10^6$ s$^{-1}$}, $k_y \ell_{e0} = 1.167$  and  $\omega_r = 27.1$ MHz.

The effects of a nonzero (moderate) $k_z$ on the MIA modes are: introducing an ion-Doppler shift $k_z u_{zi0}$ in the frequencies, shifting the unstable bands in $k_y$, and changing mildly the growth rates 
(find a more exhaustive analysis in \cite{ducr06}).

To reduce the computational cost, simulations here correspond to a hydrogen plasma.
The frequencies for hydrogen are, approximately, one order of magnitude higher than those expected in xenon.
This is a reasonable result, since equation \eqref{eq: ECDI} shows that frequency and growth of the ECDI modes are proportional to $k c_{s0}$ and, thus, scale with $1/\sqrt{m_i}$.
This trend is, indeed, retrieved in PIC simulations shown later in this paper. 
The use of hydrogen instead of xenon in PIC simulations of the ECDI is a computational advantage since it allows us to 
observe the same physical phenomena but in a shorter time. 
With the same time step (still limited by the electron dynamics), this means reducing the number of time steps by one order of magnitude.

\section{The classical ECDI: nonlinear evolution}
\label{sec: periodic}

\subsection{The numerical PIC model}
\label{sec: numerical model}

\begin{figure}[t]
    \centering
    \includegraphics[width=0.5\textwidth]{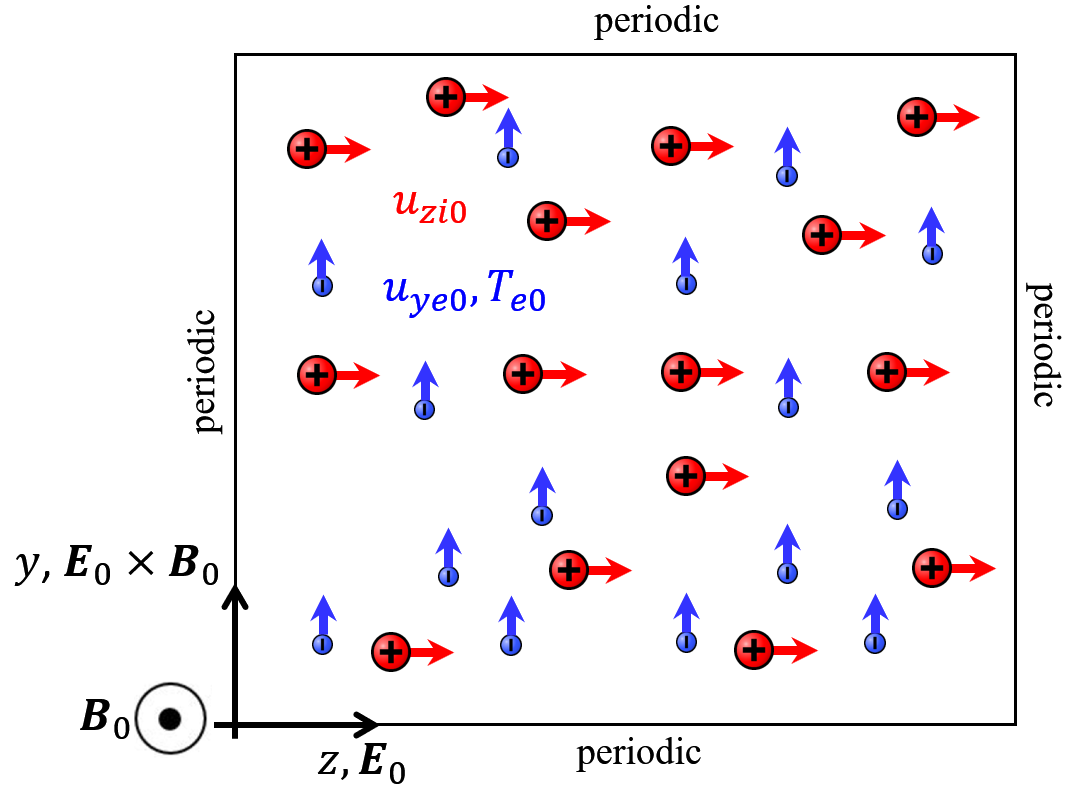}
    \caption{Diagram summarizing the simulation axes, boundary conditions and initial equilibrium state for simulations of the classical ECDI. A second type of simulations is considered that substitutes left and right boundary conditions to $\phi = 0$ and injection/absorbing conditions for particles. 
    }
    \label{fig: diagram}
\end{figure}

The nonlinear evolution and saturation of the classical ECDI is studied with a 2D axial azimuthal, 2D($z, \theta$), full PIC code developed in-house \eb{and called PICASO}.
The PIC formulation follows ions and electrons. The electrostatic potential is obtained from a Poisson solver.
The numerical codes are implemented in Fortran and use OpenMP shared-memory parallel computing.

The PIC code applies a standard Boris method to move 
electron and ion macroparticles in the periodic domain, and it employs a particle-decomposition strategy for parallel calculations.
Macroparticles have  equal and constant weights (i.e., number of real particles per macroparticle).
As already pointed out, collisions between particles are totally disregarded.

The Poisson solver is able to use different schemes depending on boundary conditions. When all boundaries are periodic (the case in the present section \ref{sec: periodic}), spectral methods are well suited to solve the Poisson equation in the Fourier complex space; here, the FFTW3 library \cite{frig05} for Fourier and inverse transform operations is used, with  a zero average potential.
If Neumann or Dirichlet conditions are used in at least one boundary (the case in section \ref{sec: injection}), the Poisson solver uses a second order finite difference scheme for the Laplace operator and electric field, and the discrete linear system is solved with the PARDISO direct-solver routines in the Math Kernel Library of INTEL.

\begin{figure}[!t]
    \centering
    \includegraphics[width=0.47\textwidth]{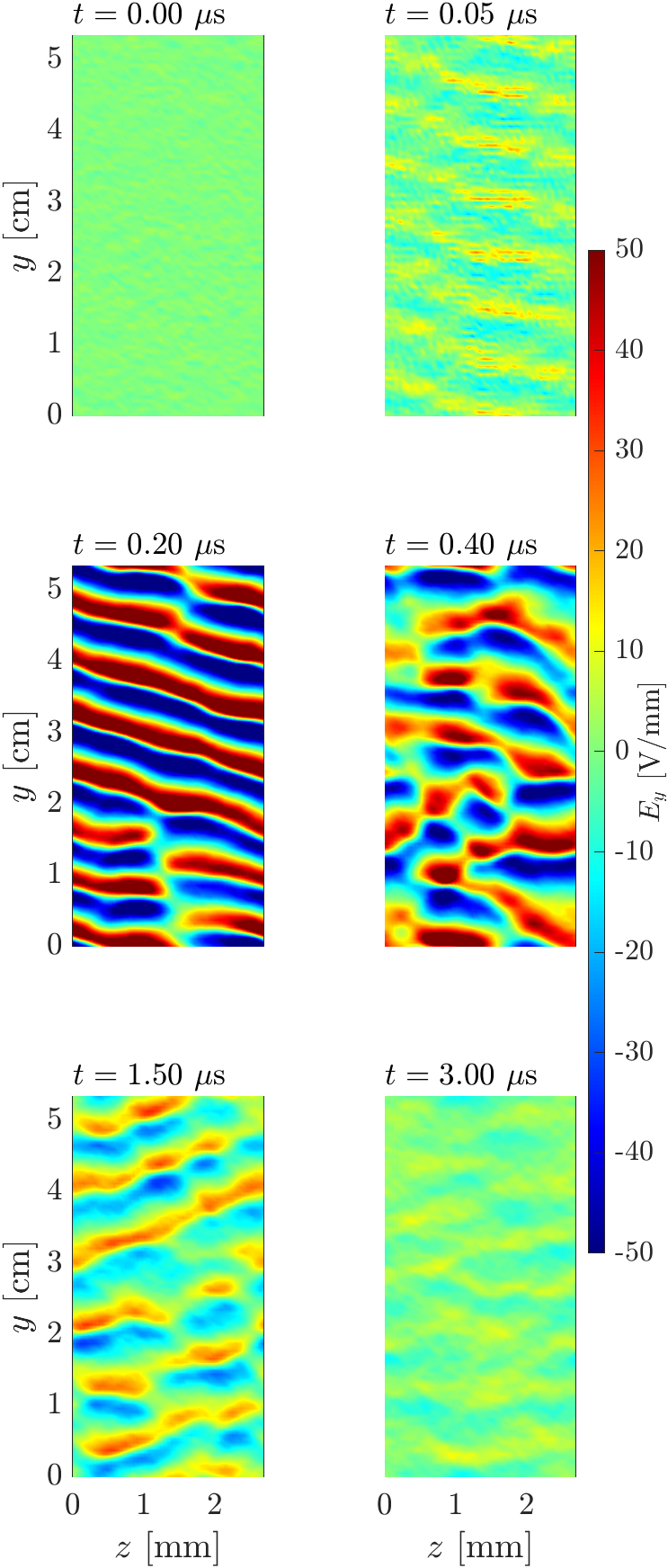}
    \caption{Reference case with fully periodic conditions: time-evolution of $E_y(y,z,t)$ in the  $yz$-plane.
    }
    \label{fig: ECDI ref Ey zy}
\end{figure}

\begin{figure}[!t]
    \centering
	\begin{subfigure}[b]{0.49\textwidth}
		\caption{$L_y = 2.679 ~ \textmd{mm}$}
		\centering
        \includegraphics[width=0.95\textwidth]{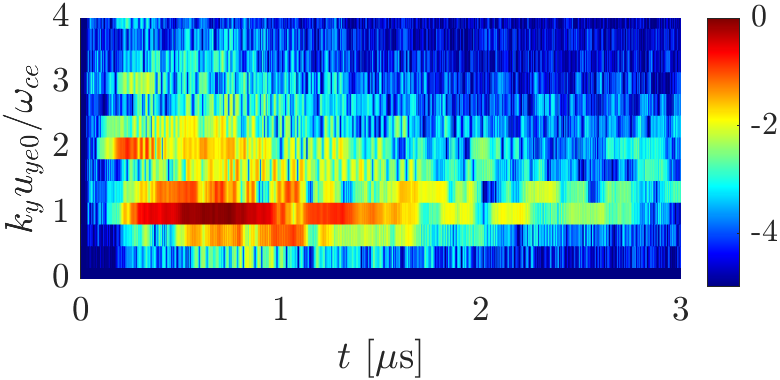}
	\end{subfigure}
    \par\bigskip
	\begin{subfigure}[b]{0.49\textwidth}
		\caption{$L_y = 5.359 ~ \textmd{mm}$}
		\centering
        \includegraphics[width=0.95\textwidth]{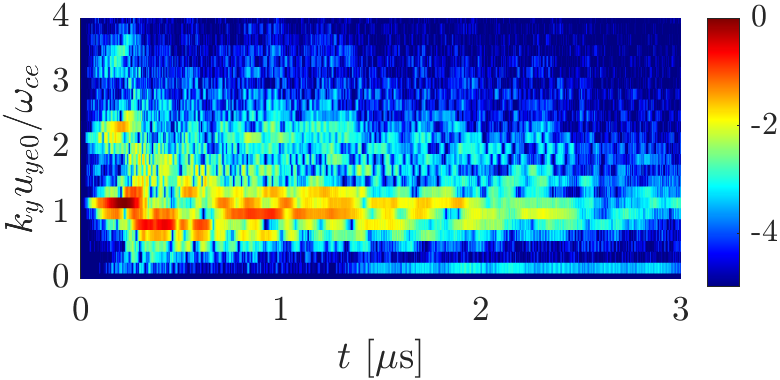}
	\end{subfigure}
    \par\bigskip
	\begin{subfigure}[b]{0.49\textwidth}
		\caption{$L_y = 10.72 ~ \textmd{mm}$}
		\centering
        \includegraphics[width=0.95\textwidth]{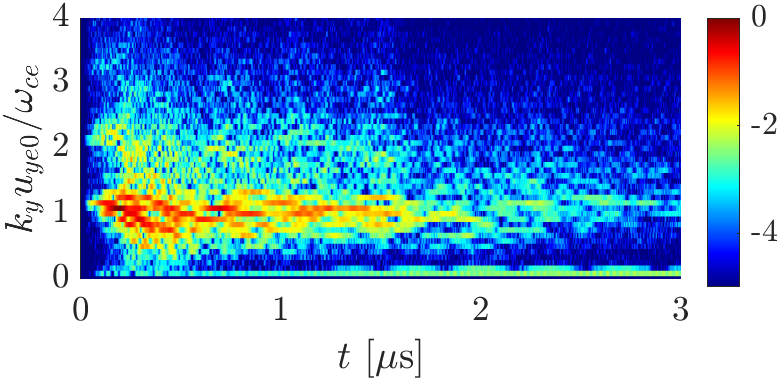}
	\end{subfigure}
    \par\bigskip
	\begin{subfigure}[b]{0.49\textwidth}
		\caption{$L_y = 16.08 ~ \textmd{mm}$}
		\centering
        \includegraphics[width=0.95\textwidth]{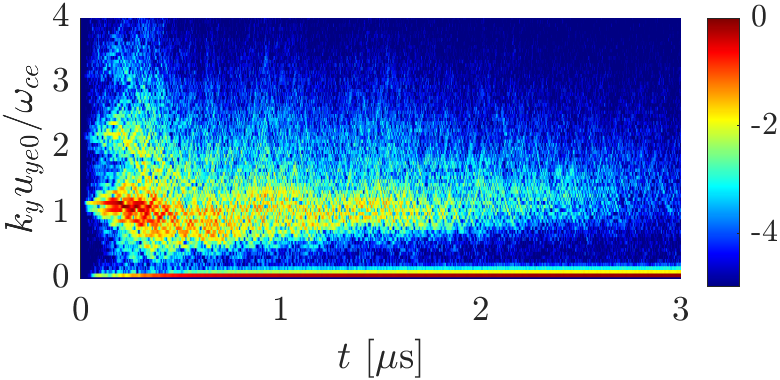}
	\end{subfigure}
    \caption{Reference case with fully periodic conditions: time-evolution of the logarithm of normalized coefficients from the fast Fourier transform of $E_y(t,y,z^*)$ in coordinate $y$, at  $z^* = 5 L_z/6$, for different $L_y$.
    }
    \label{fig: ECDI ref fft 1D y}
\end{figure}

\begin{figure}[!t]
	\centering
	\begin{subfigure}[b]{0.37\textwidth}
		\caption{$0 ~ \mu  \textmd s < t < 0.20 ~ \mu \textmd s $}
		\centering
		\includegraphics[width=\textwidth]{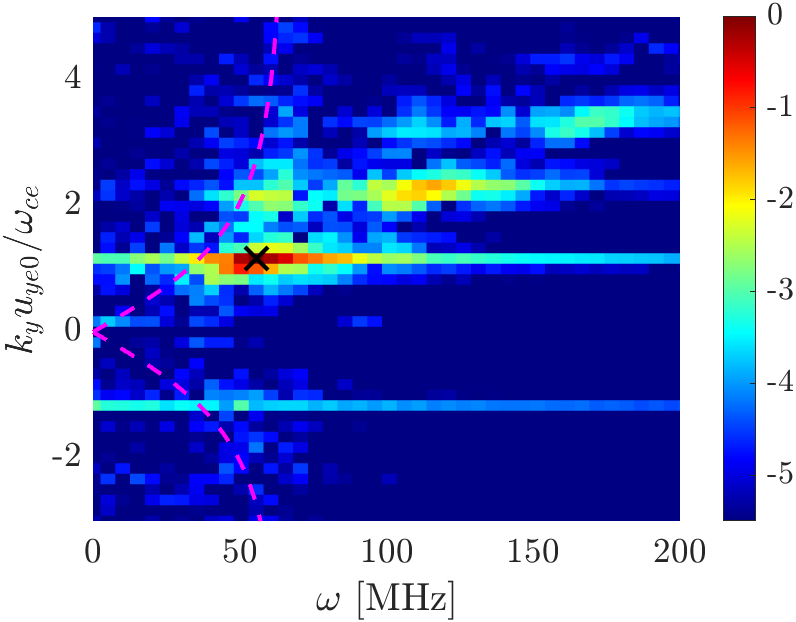}
	\end{subfigure}
    \par\bigskip
	\begin{subfigure}[b]{0.37\textwidth}
		\caption{$0.20 ~ \mu  \textmd s < t < 0.75 ~ \mu \textmd s $}
		\centering
		\includegraphics[width=\textwidth]{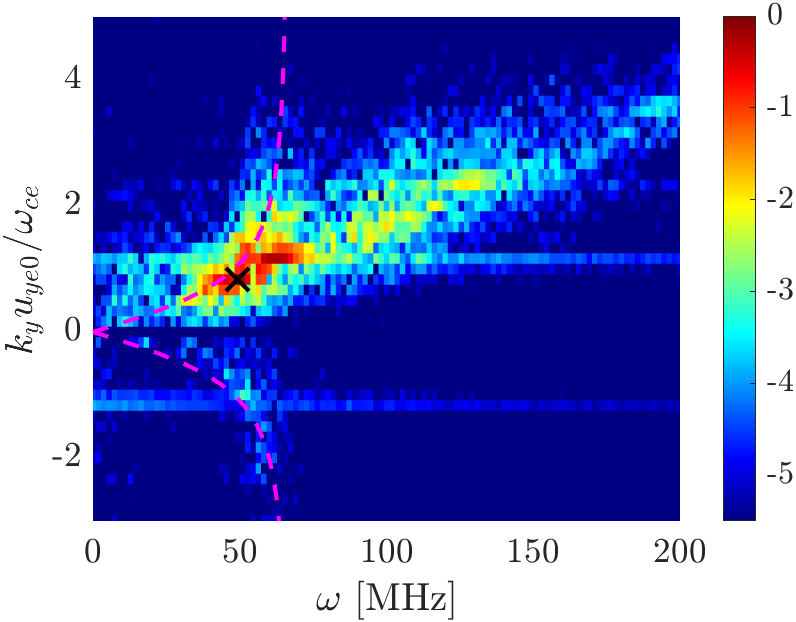}
	\end{subfigure}
    \par\bigskip
	\begin{subfigure}[b]{0.37\textwidth}
		\caption{$2.5 ~ \mu  \textmd s < t < 3.0 ~ \mu \textmd s $}
		\centering
		\includegraphics[width=\textwidth]{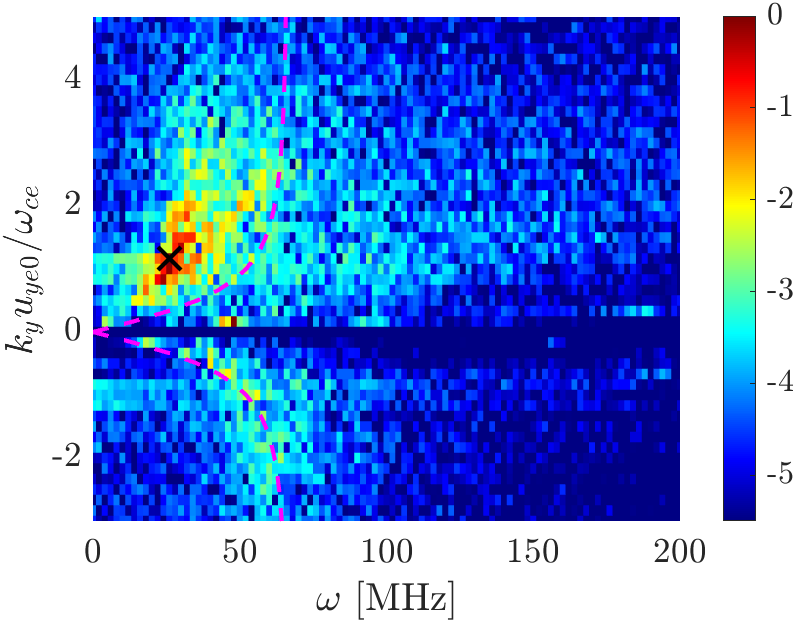}
	\end{subfigure}
	\caption{Reference case with fully periodic conditions: logarithm of normalized coefficients from the 2D fast Fourier transform of  $E_y(t,y,z^*)$ in coordinates $y$ and $t$, at $z^* = 5 L_z/6$, and for 3 different time windows. The absolute maxima are marked with a black cross. Dashed magenta lines stand for ion-acoustic modes  $\pm \omega_{\textmd{IA}}$ with $T_{e0}$ taken as the average value on the considered window: (a) 11.4 eV, (b) 40.1 eV and (c) 58.3 eV. } 
	\label{fig: 2D fft}
\end{figure}

The electric and magentic fields felt by each species and their initial macroscopic properties comply with the hypotheses and equilibrium state of section \ref{sec: dispersion relation}.
That is to say, the initial electron population is randomly sampled from a Maxwellian VDF with density $n_0$, velocity $u_{ye0} \bm 1_y$ and temperature $T_{e0}$.
The initial ion population is generated with $n_0$ and $u_{zi0} \bm 1_z$ without dispersion.
The values of these properties and every other physical and numerical parameter are gathered in Table \ref{tab: parameters}.
The macroparticle weights are kept constant throughout the simulation and can be computed from initial conditions as $n_0 \Delta y \Delta z/N_{\textmd{ppc}}$
Electron and ion particles are moved in a periodic domain with electric fields $\bm E_e = \bm E_0 + \bm E_1$ and $\bm E_i = \bm E_1$, respetively, being $\bm E_1 = -\nabla \phi$ the local fluctuation relative to $\bm E_0$ that comes as solution to the Poisson equation with periodic boundary conditions.
\eb{This approach is also followed by some authors \cite{smol23,janh18} and it is necessary in our 2D simulations to initially have equilibrium and prevent the development of axial inhomogeneity; 1D-azimuthal models not solving the axial direction do not capture this effect and can more freely account for the acceleration of ions due to $\bm E_0$ \cite{lafl16a, tacc19c}}.
Figure \ref{fig: diagram}(a) sketches the simulation setup.
At the initial equilibrium  state, the axial current of electrons, $j_{ze0}$ is zero. 
Any subsequent electron axial current is due to the cross-field transport generated by plasma oscillations.

Regarding the macroscopic and kinetic results to be shown, they include a moving-average during runtime on a window corresponding to $N_{\textmd{print}}$ time steps.
For an arbitrary variable $\varphi$, this is defined as
\begin{equation}
    \tilde \varphi_{k} = \frac{\tilde \varphi_{k-1} (N_{\textmd{print}}-1) + \varphi_{k}}{N_{\textmd{print}}}, 
\end{equation}
being $k$ the time step index, $\varphi_k$ the instantaneous value of  and $\tilde \varphi_k$  its time-average value.

\subsection{Onset, saturation, and vanishing of the ECDI}
\label{sec: onset ECDI}

The time evolution in the $yz$-plane of $E_y(t,y,z)$ is represented in figure \ref{fig: ECDI ref Ey zy}. 
The initial equilibrium state is unstable and the ECDI start to grow 
from any perturbation.
The time-dependent solution is almost 1D in the azimuthal direction, although an axial component is present in early times, mainly.
The oscillation amplitude gets a maximum  around 0.2~$\mu$s and decreases in later times until eventually  a new equilibrium state, different from the initial one, is reached. 
For $t < 0.2 ~ \mu$s, wave modes are weakly mixed and the dominant monochromatic waves are easier to observe. 
The top panels of figure \ref{fig: ECDI ref Ey zy} indicate that the dominant mode is  $n=7$,  which is the  closest one to the resonance $m = 1$.
For $t > 0.2 ~ \mu$s, there is more mixing of modes; the bottom panels of figure \ref{fig: ECDI ref Ey zy} show  transitions to $n=5$ and 6 
as dominant modes.

The linear theory showed that mode $m=2$ has a (slightly) higher growth rate than mode $m=1$. 
In fact the Fourier analysis of $E_y(t,y,z)$ in $y$ [depicted in figure \ref{fig: ECDI ref fft 1D y}(b) for our simulation with $L_y=5.359$mm]
shows some contribution of mode $m=2$ to the early time spectrum. However, the fast growth of several modes makes nonlinear effects  important soon in the simulation, which, together with the noise intrinsic to the PIC formulation, makes tough the exact comparison of early PIC results with the linear results from Vlasov equation \cite{smol23,smol21}. In addition, the long-term dominant modes in the nonlinear stage may not coincide with the most unstable modes in the linear dispersion relation \cite{janh18b}. The use of quiet start techniques by other authors \cite{smol21,lafl16a} to minimize the noise of the initial population has not been seen to be completely satisfactory.

\begin{figure}[!t]
    \centering
    \includegraphics[width=0.48\textwidth]{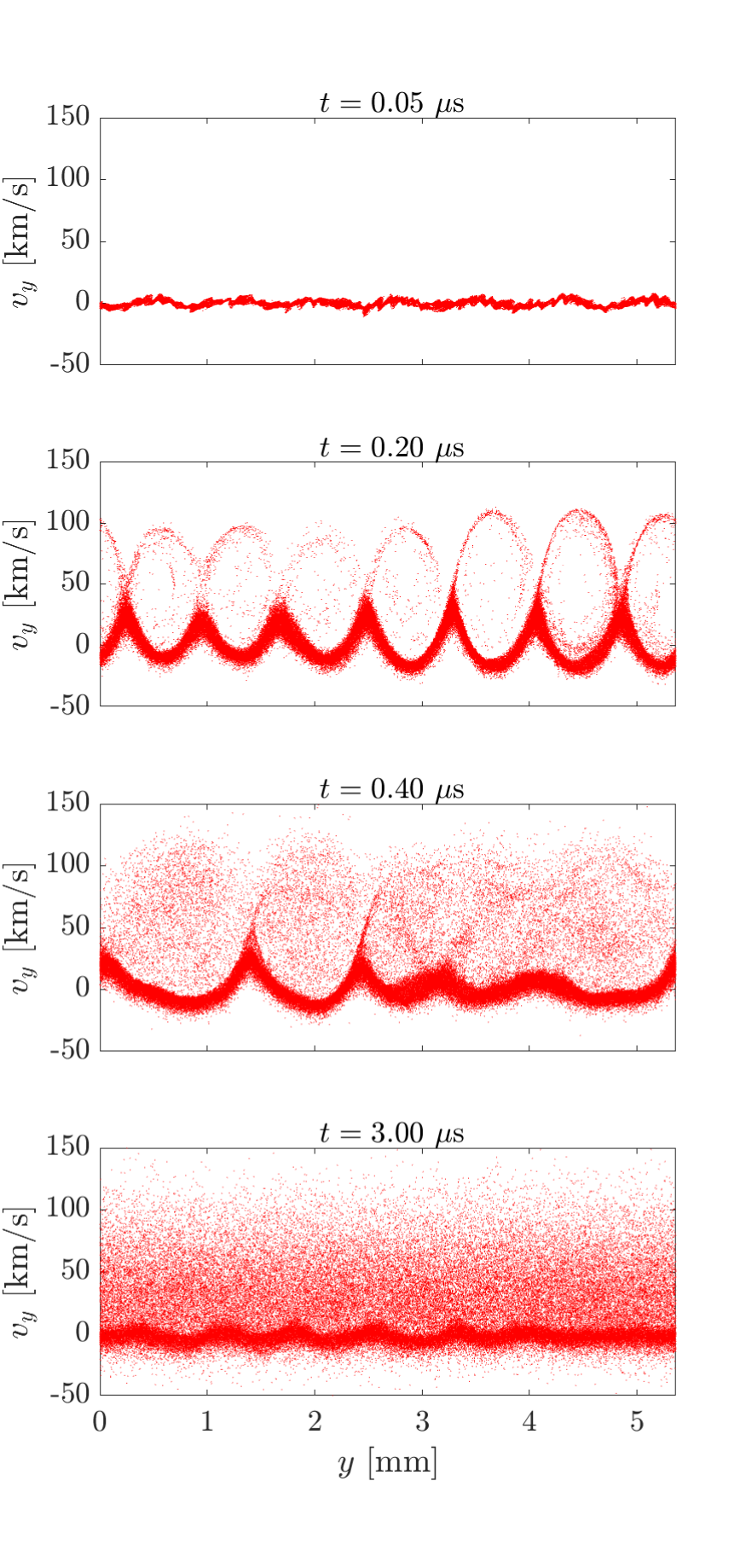}
    \caption{Reference case with fully periodic conditions: time evolution, in the phase space $(y, v_y)$, of ions contained in an axial slab of width $0.1 L_z$. The formation and blurring of the vortex-like structure characteristic of ion-wave trapping is observed.}
    \label{fig: phase space}
\end{figure}

The determination of the dominant frequency is more difficult since it depends on time itself.
In Figure \ref{fig: 2D fft}, results of the 2D fast Fourier transform in $y$ and $t$ are shown for three different time windows, together with the ion-acoustic curves $\pm \omega_{\textmd{IA}}$ for the average $T_e$ within the corresponding time interval.
Each window represents different stages in the evolution of the ECDI:
\begin{itemize}
    \item[(i)] $0 < t < 0.20 ~ \mu \textmd s $. The peaks seem to concentrate in bands near $k_y \ell_{e0} = m$, similarly to the theoretical dispersion relation in figure \ref{fig: dispersion relation}. The maximum Fourier coefficient is located at $k_y \ell_{e0} = 1.166 $ and $\omega_r = 55.6$ MHz, with phase speed $\omega_r/k_y = 42.6$ km/s. This is mode number $n=7$, near the resonance $m = 1$.
    There are secondary bands close to $m=2$ and $m=3$.
    The results in this early stage are qualitatively aligned with figure \ref{fig: dispersion relation}, but the dominant frequency $\omega_r$ is larger than predicted by the linear theory. 
    
    \item[(ii)]  $0.2 ~ \mu \textmd s < t < 0.75 ~ \mu \textmd s $. The bands of the upper spectrum near the resonances have been blurred and there is an approximately linear relation between $k_y$ and $\omega_r$, resembling a linear ion-acoustic relation.
    However, some parts of the upper spectrum seem to follow the non-neutral acoustic frequency  $\omega_{\textmd{IA}}$.    
    This is aligned with previous PIC simulations\cite{lafl17b, char19b} and experiments\cite{tsik09b}.
    The peak in the spectrum is at $k_y \ell_{e0} = 0.8334$; with $n = 5$,  $\omega_r = 49.1$ MHz and $\omega_r/k_y = 52.6$ km/s and matches the ion-acoustic curve.
    Let us note that the lower part of the spectrum (propagation in the $-\bm E_0 \times \bm B_0$ direction), shows some remnants of the counter-propagating ion-acoustic wave with frequency  $-\omega_{\textmd{IA}}$.
    
    \item[(iii)]  $2.5 ~ \mu  \textmd s < t < 3 ~ \mu \textmd s $. Even if a peak is identifiable at $k_y \ell_{e0} = 1.166$ ($n=7$), the mixing of different temporal and azimuthal scales  result in a messy spectrum without a clear dominant mode. 
    The ion-acoustic behavior is only seen in the lower part of the spectrum.   
\end{itemize}

\begin{figure}[!t]
    \centering
    \includegraphics[width=0.48\textwidth]{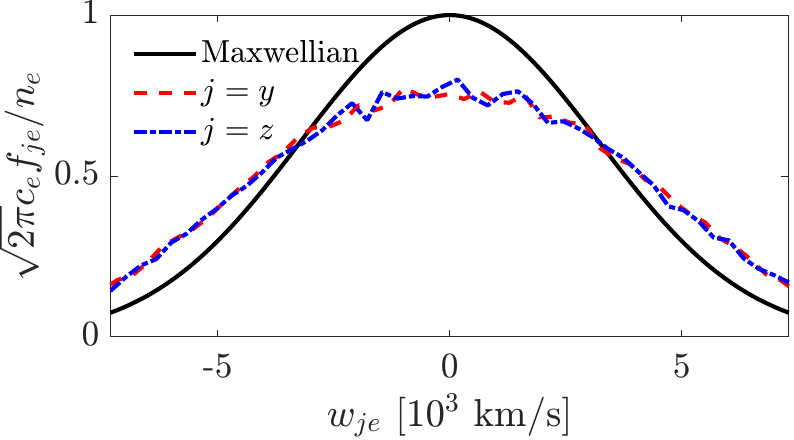}
    \caption{Reference case with fully periodic conditions: normalized electron distribution functions $f_{je}$ in the azimuthal ($j=y$, red dashed) and axial ($j=z$, blue dash-dotted) directions, being the abscissa coordinate $w_{je}$ the random velocity of electrons . The Maxwellian curve (black solid) is included for comparison, using local final values of $n_e = 10^{17}$ and $T_e = 58.8$ eV from the PIC simulation.}
    \label{fig: VDF electrons}
\end{figure}

At $t \sim  3 ~ \mu \textmd s$, the plasma seems to tend to a new equilibrium with distorted ion and electron VDFs.
Figure \ref{fig: phase space} plots
the evolution of ion particles, contained within an axial slab of width $0.1 L_z$, in phase space $(y,v_y)$.
During the growth and saturation of the ECDI (this is $t < 0.2~\mu$s), vortexes are formed in phase space showing the characteristic behavior of ion trapping in the electrostatic wave. 
The vortexes are shifted towards positive velocities, matching the $\bm E_0 \times \bm B_0$  and dominant mode propagation directions.
Later times reveal the  distortion of those vortex structures until they are fully blurred into a strongly one-sided distribution with a long tail into positive velocities.
This process coincides with the quenching of $E_y$ oscillations.
Electrons one-dimensional VDFs at the end of the simulation are shown in figure \ref{fig: VDF electrons} to be fairly isotropic and flatter close to the average velocity compared to a Maxwellian, which coincides with \cite{janh18, ducr06}.

\subsection{Evolution of the plasma energy}

\begin{figure}[!t]
	\centering
	\begin{subfigure}[b]{0.45\textwidth}
		\caption{Energy}
		\centering
		\includegraphics[width=0.9\textwidth]{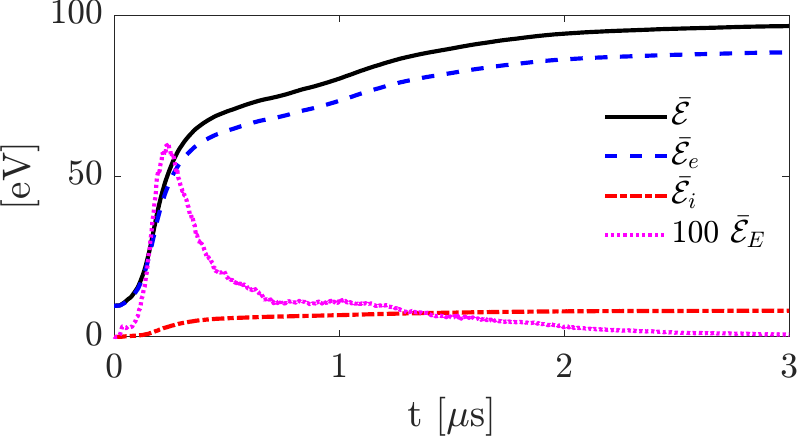}
	\end{subfigure}
    \par\bigskip
	\begin{subfigure}[b]{0.45\textwidth}
		\caption{Energy derivative}
		\centering
		\includegraphics[width=0.9\textwidth]{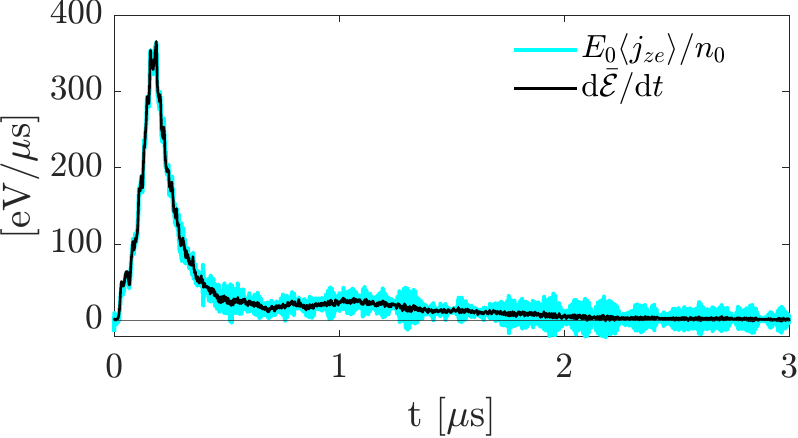}
	\end{subfigure}
    \par\bigskip
	\begin{subfigure}[b]{0.45\textwidth}
		\centering
            \caption{Effect of $L_y$ on $\bar{\mathcal E}$}
		\includegraphics[width=0.9\textwidth]{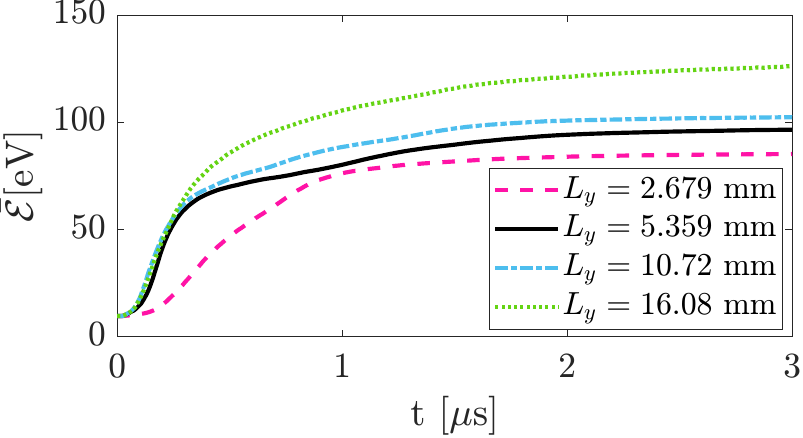}
	\end{subfigure}
    \par\bigskip
	\begin{subfigure}[b]{0.45\textwidth}
		\centering
            \caption{Effect of $L_y$ on $\angl{j_{ze}}$}
		\includegraphics[width=0.9\textwidth]{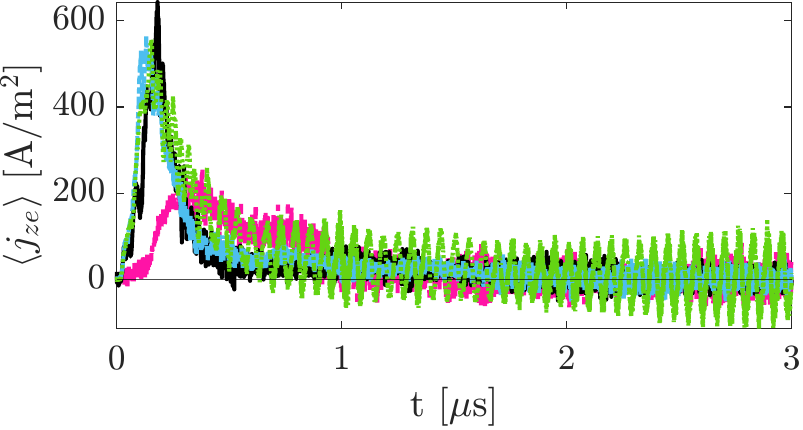}
	\end{subfigure}    
	\caption{Fully-periodic case. Time evolution of (a) the energy time-derivative and (b) the electrostatic, species and total energies per particle (i.e., $\bar{\mathcal E} = \mathcal E / \angl{n} V $.
    Panels (c) and (d) plot energy per particle and volume-averaged electron axial current for different $L_y$.
 }
	\label{fig: energy}
\end{figure}

Let us get a further insight on the ECDI by analyzing the plasma energy stored in the plasma.
The total energy in the plasma domain has contributions from the electrostatic field, electrons and ions, 
according to
\begin{equation}
    \mathcal E = \mathcal E_E +\mathcal E_e + \mathcal E_i,
    \label{eq: total energy}
\end{equation}
with
\begin{equation}
    \mathcal E_E = \frac{\varepsilon_0}{2}\int_\mathcal{V}   E_1^2  ~ \text d \mathcal{V}.
\end{equation}
the  energy of the electromagnetic oscillations,  
\begin{equation}
    \mathcal E_{s} =  \int_\mathcal{V} \brkt{   \frac{1}{2} m_{s} n_{s} u_{s}^2   + \frac{3}{2} p_{s}} \text d \mathcal{V}, \quad s=e,i
    \label{eq: energy species}
\end{equation} 
the total energy of electrons and ions;
and $\mathcal{V}$ is the volume of the domain.
For convenience, the equilibrium field $\bm E_0$ has been omitted from the definition of $\mathcal E_E$, since it would only introduce a constant offset in the energy.
The energy of each species is approximated in the PIC formulation as
\begin{equation}
    \mathcal E_{s} = \sum_p  \frac{1}{2} m_s v_p^2 W_p,
    \label{eq: energy species particles}
\end{equation} 
where the sum is on every particle in the domain with $v_p$ and $W_p$ the particle speed and weight, respectively.

In a consistent situation, the work done by the electric field should act as a mechanism that converts species energy on electric-field energy, and the other way around. 
However, the assumptions behind the linear theory of the ECDI forced us to let ions and electrons feel different electric fields.
Because of this non-conventional feature, it can be proved that the total energy changes according to
\begin{equation}
    \derpar{\mathcal E}{t}
    = \int_\mathcal{V}  (\bm j_e \cdot \bm E_0) ~ \text d\mathcal{V}
    = E_0 \int_\mathcal{V} j_{ze}  ~ \text d\mathcal{V}
    = E_0 \langle j_{ze} \rangle \mathcal{V},
    \label{eq: derivative total energy}
\end{equation}
with $\bm j_e$ the electron current density and $\langle j_{ze} \rangle$ the volume-averaged $j_{ze}$.
Therefore, this kind of simulation will not show a proper conservation of energy and the equilibrium electric field will pump energy into the isolated system.
However, this source of energy requires also an axial electron current $j_{ze}$ to be developed. 
This means that the energy is conserved initially until the instability is triggered, and any other stationary energy state should satisfy a null $\langle j_{ze} \rangle$, i.e., no turbulent electron transport.

Figure \ref{fig: energy}(a) shows the evolution of electrostatic, species, and total energies in the domain per real particle. 
Here, average energies per particle $\bar{\mathcal E} = \mathcal E/\angl{n} \mathcal V$ are used, being $\angl{n}$ the volume-averaged density and $\angl{n}  \mathcal V$ the real number particles in the domain.
Initially $\bar{\mathcal E}_E = 0$, $\bar{\mathcal E}_i = m_i u_{zi0}^2/2 = 0.0324$ eV and $\bar{\mathcal E}_e = 3T_{e0}/2 +  m_e u_{ye0}^2/2 = 9.71$ eV.
Eventually, the total energy saturates and becomes stationary. There is significant heating of electrons and, to a lesser extent, of ions; which is coherent with other works \cite{smol23,janh18,lafl16a,tacc19c}. The electrostatic energy is much lower than those of electrons and ions, and approaches zero for late simulation times, when oscillations are very weak.

The balance \eqref{eq: derivative total energy} is evaluated in figure \ref{fig: energy}(b).
The rate $\dertotinline{\bar{\mathcal E}}{t}$ is approximated numerically and compared with the source term $E_0 \langle j_{ze} \rangle / n_0$.
The two curves show an excellent match apart from the noise inherent to the PIC approach, specially problematic in the calculation of $j_{ze}$. The PIC simulation is able to approximately replicate the theoretical energy balance \eqref{eq: derivative total energy}, which is a good sign of the validity of such simulations. 
The change in energy shows a peak and then decreases tending to zero for late simulation times, meaning that a new stationary equilibrium state of the plasma is reached.  
As already said, the energy source is related with the development of a net axial current, so the new equilibrium holds an average $j_{ze}$ equal to zero. 
However, at mid simulation times, saturated ECDI modes are effective in inducing an axial transport of electrons.

\subsection{Effect of the domain's azimuthal length}

As pointed out in section \ref{sec: dispersion relation}, the azimuthal domain length $L_y$ is an important parameter that determines the unstable ECDI modes that can be excited in a finite domain. 
On the contrary, the effect of $L_z$ has been seen to be small, although the chosen axial length is large enough to allow the formation of waves with axial component fitting $L_z$.
Increasing $L_y$ allows larger scales to develop and increases the spectral resolution in the $k_y$-space, better capturing the continuous dispersion relation for an infinite plasma. In figure \ref{fig: ECDI ref fft 1D y}, the time evolution of azimuthal Fourier coefficients for $E_{y1}$ are represented for several $L_y$ multiples of 2.679 mm, the reference case corresponding to plot \ref{fig: ECDI ref fft 1D y}(b).
The main trends identified in the shortest case are also recovered in larger domains: modes close to resonant bands $m=1$ and $m=2$ are excited in early times, modes close to $m=1$ dominate at mid times, and oscillations quench after the spectrum peaks are passed. These trends become clearer as longer domains are used. Some differences worth mentioning among cases are: (i) a mode with  azimuthal wavelength equal to $L_y$ (i.e. $n = 1$) 
is present only at long $L_y$
and (ii) the enhanced spectral resolution of longer $L_y$ allows to better capture shorter resonant bands, such as the modes close to $m=2$.
Similar conclusions on these two points are reached in reference \cite{smol23}.

 Figure \ref{fig: energy}(c) shows that the energy per real particle increases with parameter $L_y$, probably due to the greater number of unstable excited modes \cite{smol23}.
 With the exception of the shortest case $L_y = 2.679$ mm, which shows a slower energy increase and lower $\langle j_{ze} \rangle$  [most probably due to a poor spectral resolution, unable to properly capture the peaks of the growth rate of Fig \ref{fig: dispersion relation}] 
 the rates of energy increase during the growth period are similar for $L_y \geq 5.359$ mm.
 This suggests that the maximum $\langle j_{ze} \rangle$ and the time to reach it are numerically robust and thus physical.
 Therefore, results with  $L_y = 5.359$ mm are representative of simulations with longer $L_y$ multiples of 5.359 mm.
 This behavior disagrees with reference \cite{smol23}, where changes in $L_y$ can 
 drastically change the transients of $\langle j_{ze} \rangle$ and, thus, the energy growth rates.

 Since the long term behavior of $\langle j_{ze} \rangle$ tends to zero in every case, we can conclude that the long-wavelength mode with $n=1$ that develops with increasing $L_y$ is not effectively producing an axial electron transport. 
 This is consistent with the late evolution of $n_e$ and $E_y$ (not included), which shows completely out of phase oscillations.

\subsection{Effect of the ion mass}
\label{sec: ion mass}

\begin{figure}[!t]
    \centering
	\begin{subfigure}[b]{0.49\textwidth}
		\caption{$m_i = 1 \textmd{u}$}
		\centering
        \includegraphics[width=\textwidth]{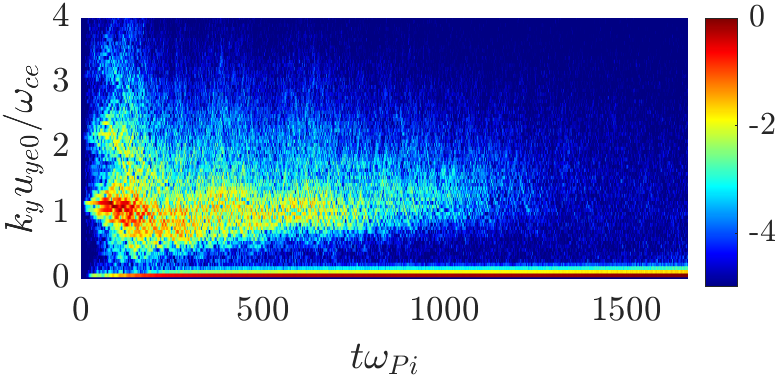}
	\end{subfigure}
    \par\bigskip
	\begin{subfigure}[b]{0.49\textwidth}
		\caption{$m_i = 4  \textmd{u}$}
		\centering
        \includegraphics[width=\textwidth]{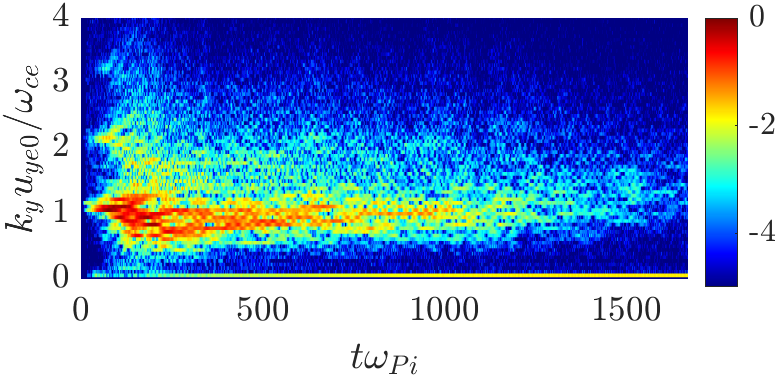}
	\end{subfigure}
    \par\bigskip
	\begin{subfigure}[b]{0.49\textwidth}
		\caption{$m_i = 9  \textmd{u}$}
		\centering
        \includegraphics[width=\textwidth]{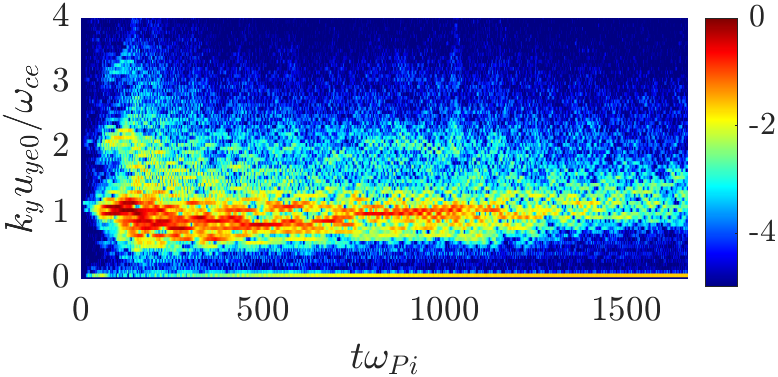}
	\end{subfigure}
    \caption{Fully-periodic case, Time-evolution of the logarithm of normalized azimuthal Fourier coefficients for different values of $m_i$.
    }
    \label{fig: ECDI mi fft 1D y}
\end{figure}

\begin{figure}[!t ]
    \centering
    \includegraphics[width=0.5\textwidth]{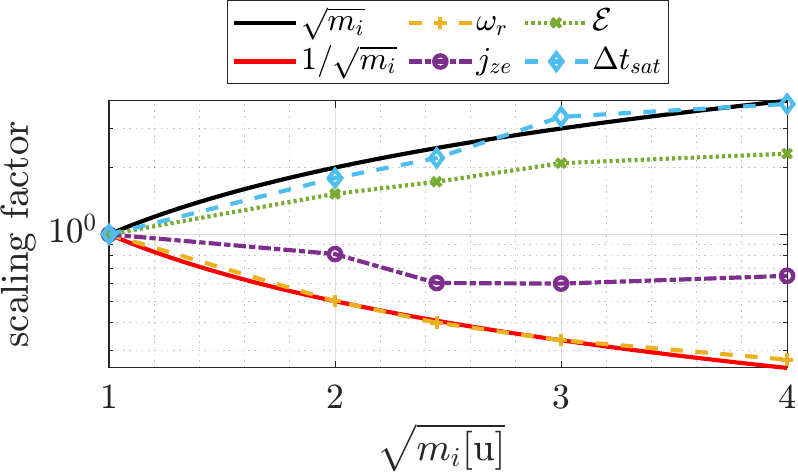}
    \caption{Scaling factor, with the ion mass $m_i$, of several variables: dominant $\omega_r$ in the interval $t \omega_{Pi} < 83.15$, peak averaged $j_{ze}$, final $\mathcal E$ and $\Delta t_{\textmd{sat}}$. The curves $\sqrt{m_i}$ and $1/\sqrt{m_i}$ are also included for comparison. }
    \label{fig: scaling mi}
\end{figure}

In this subsection, the simulation is repeated for increasing values of the ion mass. While hydrogen mass, $m_H$ has been used in order to speed up the dynamics and minimize the computational cost, the practical cases of interest deal with heavier ions, e.g., xenon or krypton. 
In addition, an increased azimuthal length $L_y = 16.08$ mm is used to have a proper spectral resolution and a fair comparison, since an increased $m_i$ narrows the unstable bands in the dispersion relation near the resonances what needs an increased spectral resolution. 

In the early evolution of the ECDI, the dispersion relation \eqref{eq: ECDI} shows that the frequency should scale inversely proportional to $\sqrt{m_i}$. However, this does not have to be the case of the nonlinear dynamics or other magnitudes.
The scaling factors for several quantities, defined as the ratio with respect to the value in the reference case $m_i = 1$ u
are computed in figure \ref{fig: scaling mi}: dominant $\omega_r$ in the interval $t \omega_{Pi} < 83.15$, peak of $\angl{j_{ze}}$, final $\mathcal E$ and $\Delta t_{\textmd{sat}}$. 
The observed trends are coherent with results reported in reference \cite{chen23}.
If the dominant frequency is measured in early simulation times, the observed scaling perfectly matches the linear theoretical result.
The saturation time $\Delta t_{\textmd{sat}}$, defined as the time of maximum average $\angl{j_{ze}}$, seems to scale proportional to $\sqrt{m_i}$. 
The 1D azimuthal Fourier spectra are compared in figure \ref{fig: ECDI mi fft 1D y}, where the time axis is normalized with the inverse of the ion plasma frequency $\omega_{Pi}$. 
While the saturation of the early oscillations produced by the ECDI happens at similar $t \omega_{Pi}$ (as shown by the scaling of  $\Delta t_{\textmd{sat}}$), the quenching of short wavelength $m=1$ and $\angl{j_{ze}}$ seems to be slightly slower with increasing $m_i$. 

The level of turbulent current, measured by the peak $\angl{j_{ze}}$, initially decreases with $m_i$, which could be explained by the weaker oscillations observed in $n_e$ (not included here), but seems to saturate. The current $\angl{j_{ze}}$ decreases, however, slower than $1/\sqrt{m_i}$ which leads to a final $\mathcal E$ increasing with $m_i$, in accordance with equation \eqref{eq: derivative total energy}.
Extrapolating these results, with xenon we can expect somehow smaller level of transport but dilated in time, what would cause a significant heating such as that observed in similar studies in the literature \cite{lafl16a,janh18b,smol23}.

\subsection{Saturation behavior in previous literature}

\begin{figure*}[!t]
    \centering
    \includegraphics[width=0.8\textwidth]{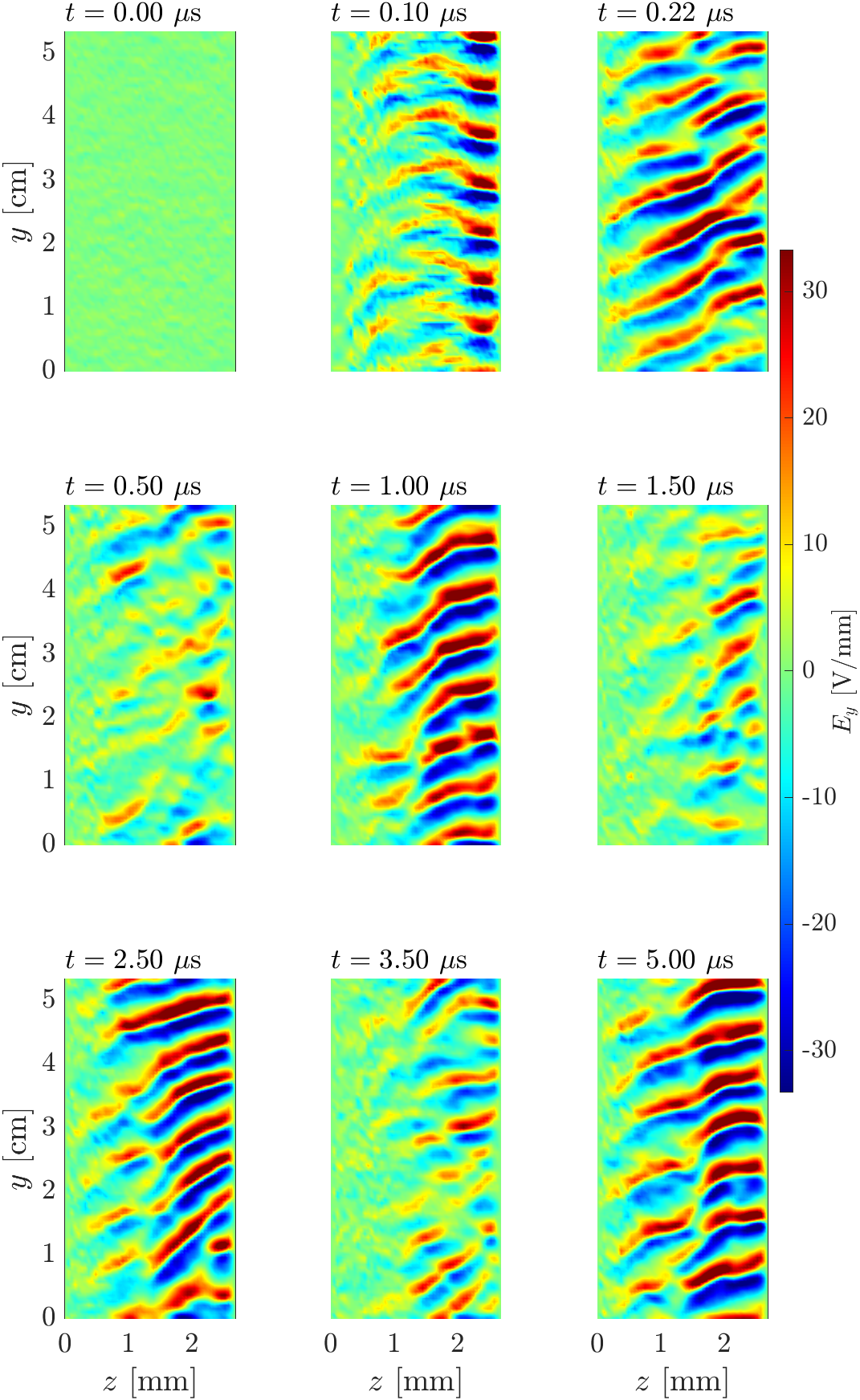}
    \caption{Finite plasma with axial injection.
    Time-evolution of $E_y$ in the  $yz$-plane.
    }
    \label{fig: ECDI inj Ey zy}
\end{figure*}

The results we have shown in this section pursue the simulation of scenarios as close as possible to the classical ECDI.
While we observe the growth and saturation of oscillations due to an instability, modes that induce an axial electron transport seem to vanish at long times.
\eb{While the initial behavior agrees well with previous efforts, the late simulation behavior differs from other 1D-azimuthal works \cite{janh18,lafl16a,tacc19c,smol23}, which report sustained oscillations for late simulation times that either forever grow \cite{janh18,smol23} or saturate \cite{lafl16a, tacc19c} if particle velocities are refreshed after travelling a certain virtual axial distance.}

\begin{figure*}[!t]
	\centering
	\begin{subfigure}[b]{0.28\textwidth}
		\centering
		\caption{$u_{zi0} = 1 ~ \textmd{km/s}$ }
		\includegraphics[width=0.95\textwidth]{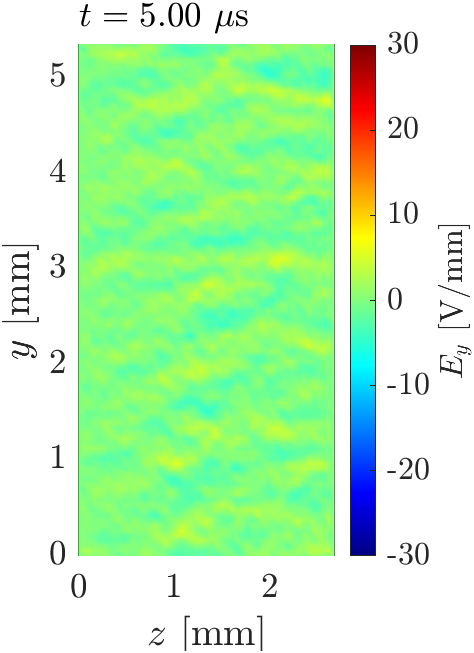}
	\end{subfigure}
	\begin{subfigure}[b]{0.28\textwidth}
		\centering
		\caption{$u_{zi0} = 10 ~ \textmd{km/s}$}
		\includegraphics[width=0.95\textwidth]{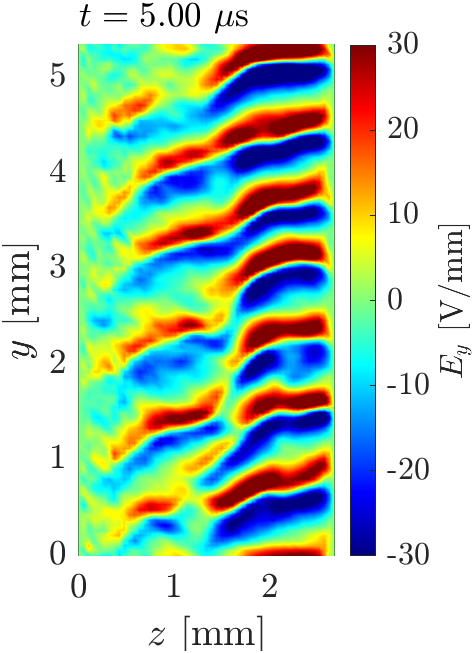}
	\end{subfigure}
	\begin{subfigure}[b]{0.28\textwidth}
		\centering
		\caption{$u_{zi0} = 100  ~ \textmd{km/s}$}
		\includegraphics[width=0.95\textwidth]{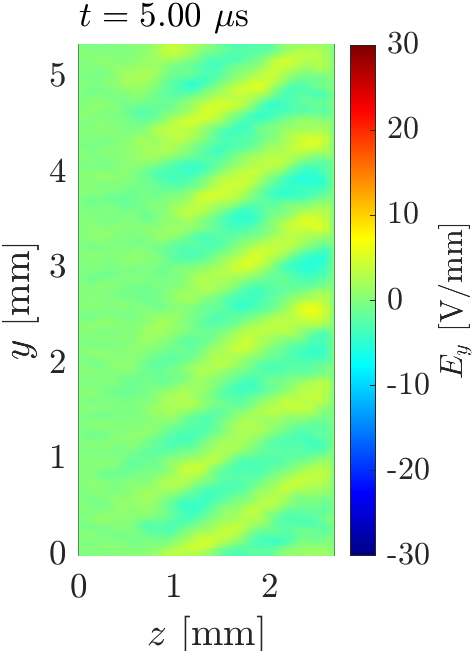}
	\end{subfigure}
	\caption{Finite plasma with axial injection. Late behavior (at $t=5 \mu$s) of $E_y$ in the  $yz$-plane, for different $u_{zi0}$.}
	\label{fig: ECDI inj Ey final}
\end{figure*}

When no virtual axial dimension or refreshing are considered, previous literature \cite{smol23,janh18,lafl16a} observe an unlimited growth of the oscillations and heating. 
The reasons that explain why they do not see the quenching of short-wavelength waves 
are still to be investigated, but we can speculate about a couple of possibilities. 
An explanation could be that the evolution towards a new equilibrium is not inherent to all simulations of this type and depends on the selection of parameters (although this is the behavior we have seen in all of our simulations). 
Another possibility, in line with the results in section \ref{sec: ion mass}, 
could be that more simulation time is needed with xenon than considered in \cite{smol23,janh18,lafl16a} to observe our late-simulation stage.


\section{THE ECDI IN AN AXIALLY-INJECTED FINITE-PLASMA} 
\label{sec: injection}

\eb{As seen in section \ref{sec: onset ECDI}, the stages observed in the evolution of oscillations are greatly related to the interaction of ions with the electrostatic wave and the distribution of ion particles in phase space.
In this sense, the treatment of ion particles (fully periodic or injection) may determine the long-term behavior of the simulations. 
When a fully-periodic domain is considered, there is no change in velocity for particles going through boundaries and, thus, the VDFs of ions (see figure \ref{fig: phase space}) and electrons change from the initial ones as a result of the ECDI interaction solely.
In our periodic results, this process ultimately leads to the quenching of oscillations and electron axial transport.}
When there is particle removal and injection through boundaries, particles that have already interacted with the wave are eventually removed from the simulation and new particles are injected having a different VDF that could possibly modify the non-linear behavior of the instability and lead to sustained oscillations and transport.

A new simulation setup is presented here that replaces axial periodic conditions by injection ones, while periodic conditions are kept in the azimuthal direction. 
Any particle leaving the domain though axial boundaries is removed from the simulation. 
A constant flux of ions $n_0 u_{zi0}$ is injected through the left boundary, with zero temperature and velocity $u_{zi0} \bm 1_z$.
Constant fluxes of electrons $\pm n_0 c_{e0} / \sqrt{2\pi}$, corresponding to a half-Maxwellian, 
are injected through left and right boundaries, whose velocities are sampled from a Maxwellian VDF with temperature $T_{e0}$ and velocity $u_{ye0} \bm 1_y$.
The injection fluxes \eb{assume that plasma at the initial equilibrium exists out of our domain and, thus, their values}
are chosen such that they match the amount of ions and electrons leaving the domain in equilibrium conditions. 
\eb{Therefore, this configuration conforms an equilibrium state equivalent to the fully periodic case.
Once the instability arises the injected fluxes do not need to coincide with fluxes leaving the domain and the number of particles and electric charge are not conserved anymore.
For that reason, axial conditions on potential are also changed from periodic to fixed potential $\phi = 0$ and the finite-difference PARDISO version of the Poisson solver is used. 
}

\eb{
The motivation behind our simulations with injection is different from that of previous 1D-azimuthal models that use re-injection (i.e., conserving particles) and refreshing \cite{lafl16a,tacc19c} to reach saturation and to limit heating; phenomena that we already observe with full-periodic conditions.
Our approach is also different; we decided to inject particles with fixed fluxes to mimic the generation of particles in a finite plasma such as the Hall discharge, which we assume independent of the plasma dynamics caused by the instability.
Moreover, 1D-azimuthal models with virtual axial length do not resolve the axial direction and, thus, the axial transport of particles and waves is only partially captured; with the velocity refreshing being equivalent to inelastic collision events \cite{tacc19c} with a rate determined by the axial length.
Our model is 2D (including the axial direction) so that injection cannot longer be reduced to a collisional phenomena and the mix of new injected particles into the bulk plasma is properly resolved.
}

\begin{figure}[!t]
	\centering
	\begin{subfigure}[b]{0.49\textwidth}
		\caption{Energy balance}
		\centering
		\includegraphics[width=0.9\textwidth]{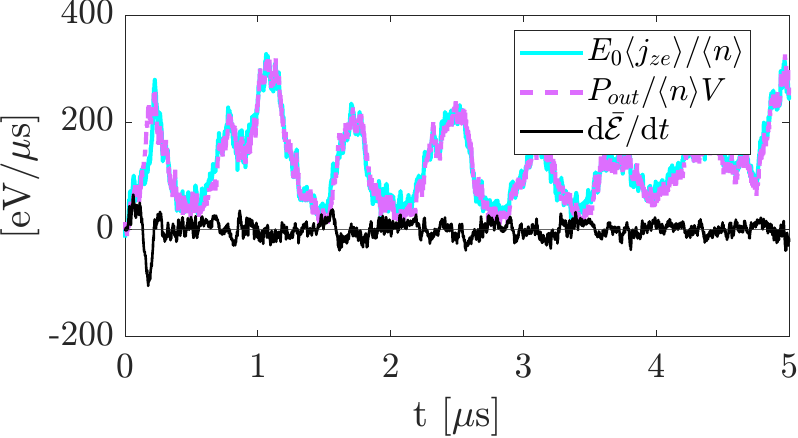}
	\end{subfigure}
	\begin{subfigure}[b]{0.49\textwidth}
		\caption{Total and species energies}
		\centering
		\includegraphics[width=0.9\textwidth]{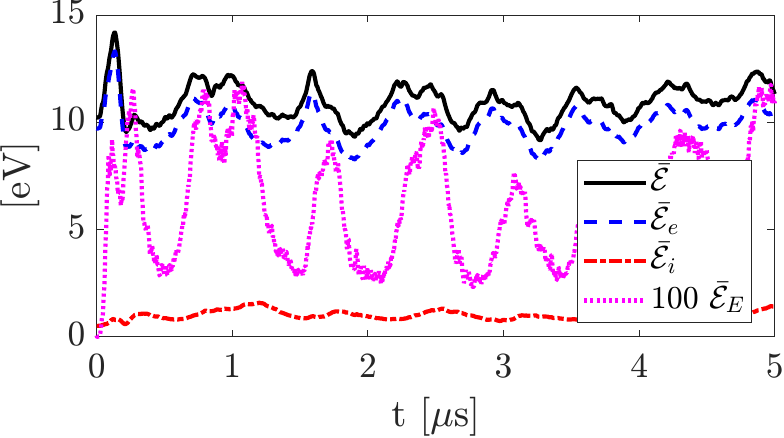}
	\end{subfigure}
	\caption{Finite plasma with axial injection. Time evolution of (a) energy balance and (b) electrostatic (c), species and total energies; for an axial-injection simulation with $u_{zi0} = 10$ km/s. Values are given per real particle.
 }
	\label{fig: energy inj}
\end{figure}

\subsection{Dependence on the ion residence time}

In section \ref{sec: periodic}, we showed the usual stages that we observe in the evolution of the ECDI, being the final one the quenching of short-wavelength oscillations. 
These stages are related with the ion velocity distribution and the formation and blurring of vortex-like structures in phase space.
In this sense, adding axial boundaries that inject and absorb particles can have a major impact on the late simulation behavior since old particles that have interacted with the wave are removed and new, non-trapped particles are injected with the original distribution.

In this scenario with axial injection, more similar to a Hall discharge, the ion residence time 
\begin{equation}
    \Delta t_{i0} = L_z/u_{zi0}
\end{equation}  
(infinite in a fully periodic domain independently of $L_z$) is the key parameter in maintaining a saturated ECDI, with sustained short-wavelength oscillations and a nonzero axial electron transport. 
Three regimes will be distinguished depending on 
$\Delta t_{i0}$ being much smaller, of the same order, or much higher than the saturation time $\Delta t_{\textmd{sat}}$ of the full-periodic configuration for the same equilibrium plasma, where $\langle j_{ze} \rangle$ peaks.

\begin{figure*}[!t]
    \centering
	\begin{subfigure}[b]{0.49\textwidth}
		\caption{$u_{zi0} = 1$ km/s}
		\centering
        \includegraphics[width=0.9\textwidth]{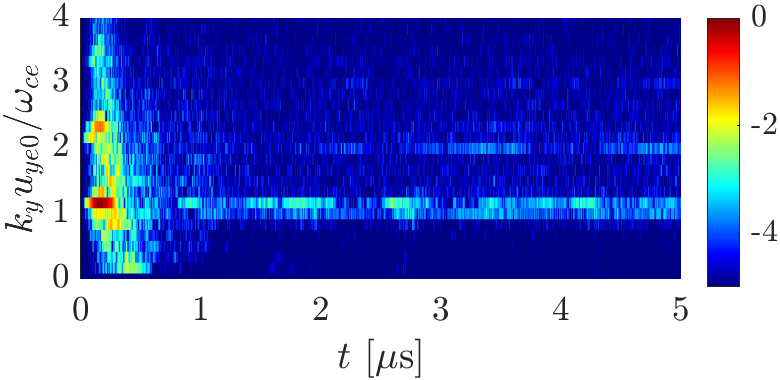}
	\end{subfigure}
	\begin{subfigure}[b]{0.49\textwidth}
		\caption{$u_{zi0} = 5$ km/s}
		\centering
        \includegraphics[width=0.9\textwidth]{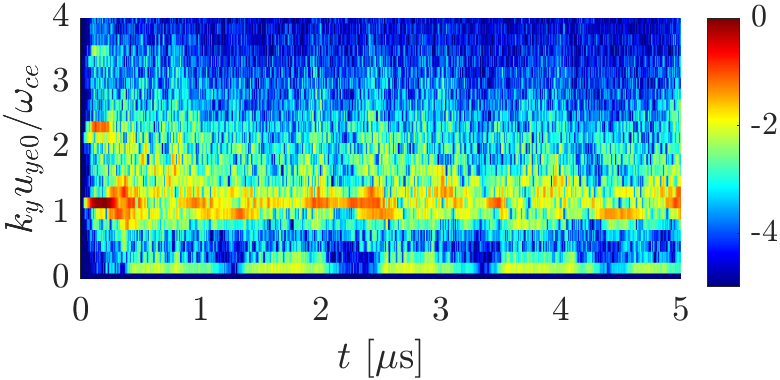}
	\end{subfigure}
    \par\bigskip
	\begin{subfigure}[b]{0.49\textwidth}
		\caption{$u_{zi0} = 10$ km/s}
		\centering
        \includegraphics[width=0.9\textwidth]{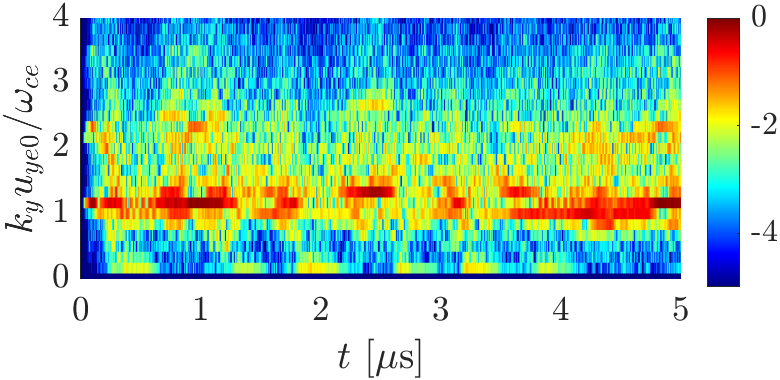}
	\end{subfigure}
	\begin{subfigure}[b]{0.49\textwidth}
		\caption{$u_{zi0} = 15$ km/s}
		\centering
        \includegraphics[width=0.9\textwidth]{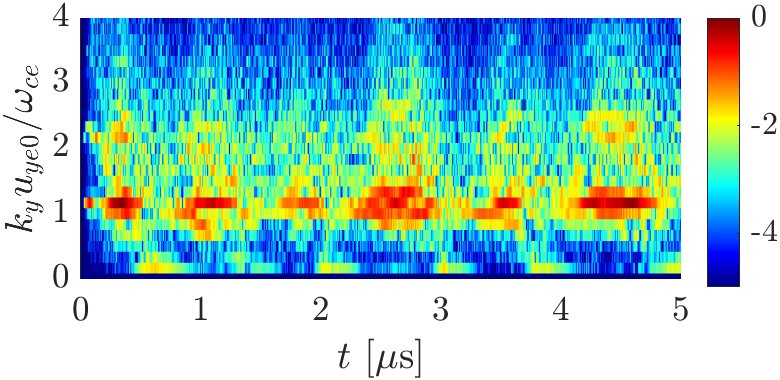}
	\end{subfigure}
    \par\bigskip
	\begin{subfigure}[b]{0.49\textwidth}
		\caption{$u_{zi0} = 25$ km/s}
		\centering
        \includegraphics[width=0.9\textwidth]{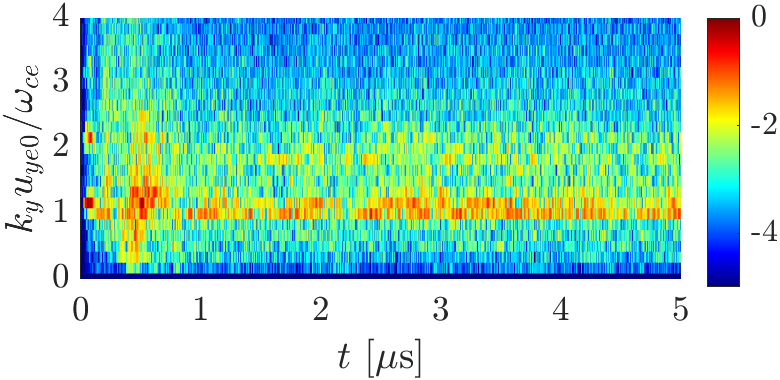}
	\end{subfigure}
	\begin{subfigure}[b]{0.49\textwidth}
		\caption{$u_{zi0} = 100$ km/s}
		\centering
        \includegraphics[width=0.9\textwidth]{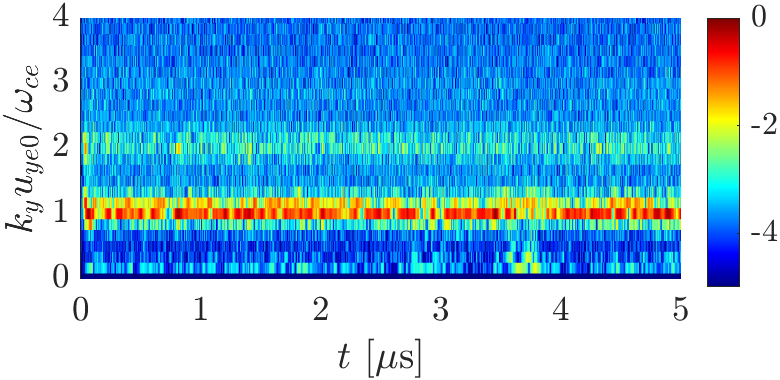}
	\end{subfigure}
    \caption{Finite plasma with axial injection. Time-evolution of the logarithm of normalized azimuthal Fourier coefficients for different values of $u_{zi0}$.
    }
    \label{fig: ECDI inj fft 1D y}
\end{figure*}

We consider again a hydrogen plasma and the reference case of Table \ref{tab: parameters}.
The saturation time 
was $\Delta t_{\textmd{sat}} \sim  0.16$-0.18 $\mu$s.
We run cases with $u_{zi0} = 1$, 5, 10, 15, 20, 25 and 100 km/s, yielding, for $L_z=2.679$ mm, residence times $\Delta t_{i0} = 2.68$, 0.536, 0.268, 0.179, 0.134, 0.107 and 0.0268 $\mu$s, covering the three expected regimes.

\begin{figure}[!t]
    \centering
	\begin{subfigure}[b]{0.49\textwidth}
		\caption{Axial electron current}
		\centering
        \includegraphics[width=\textwidth]{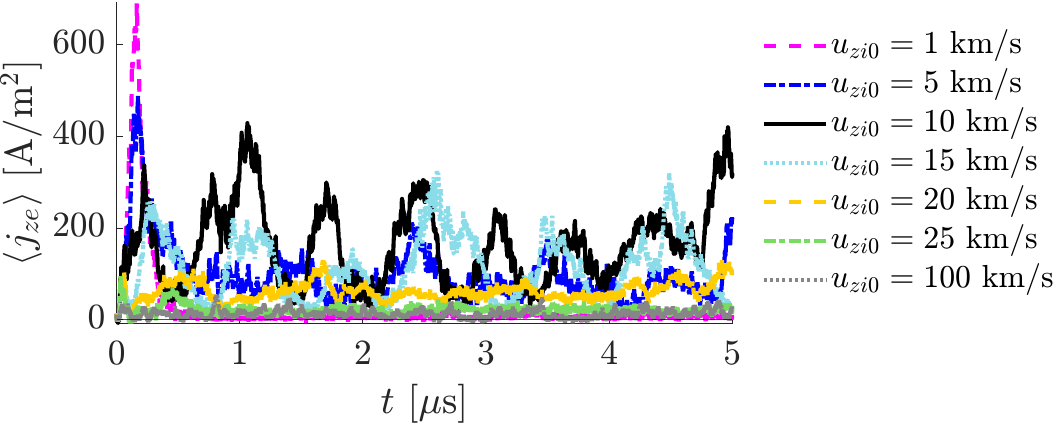}
	\end{subfigure}
    \par\bigskip
	\begin{subfigure}[b]{0.49\textwidth}
		\caption{Time-averaged $\angl{j_{ze}}$}
		\centering
        \includegraphics[width=0.95\textwidth]{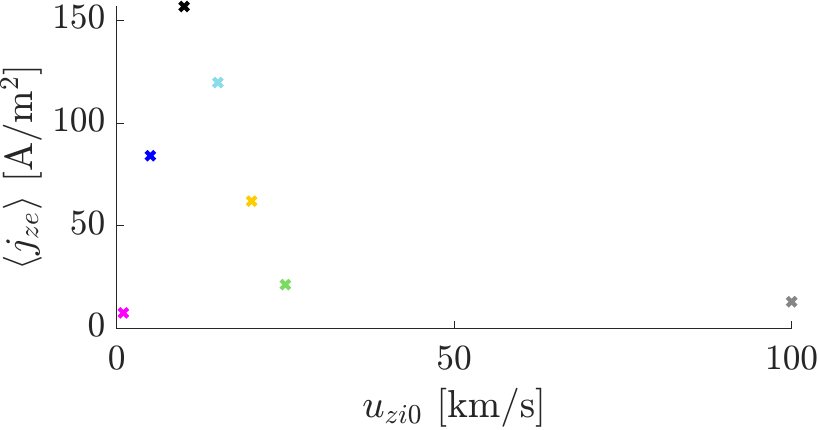}
	\end{subfigure}
     \caption{Finite plasma with axial injection. (a) Time evolution of the volume-averaged electron axial current $\angl{j_{ze}}$ for several $u_{zi0}$ and (b) its time average for $t \geq 2$ $\mu$s.
     }
    \label{fig: jze inj}
\end{figure}

The evolution of azimuthal Fourier coefficients in figure \ref{fig: ECDI inj fft 1D y} and  the average current $j_{ze}$ in figure \ref{fig: jze inj} give an idea of the time evolution, where the transition from one limit regime to the other can be appreciated.
The final $E_y$ in the axial-azimuthal plane after $5~\mu$s for cases $u_{zi0} = 1$, 10 and 100 km/s are shown in figure \ref{fig: ECDI inj Ey final}.

The case having $u_{zi0} = 1$ km/s corresponds to regime $\Delta t_{i0} \gg \Delta t_{\textmd{sat}}$ and the final $E_y$ resembles the periodic case. 
Looking at figure \ref{fig: ECDI inj fft 1D y}(a), we see that the time evolution of Fourier coefficients is only similar at very early times when the onset of the ECDI on the initial population; after which oscillations are mild. 
From the energy point of view, the different axial boundary conditions here limit the electron heating and total energy in the domain, which is probably limiting the energy  and contributing to the quenching of oscillations after the first ECDI onset. 
The evolution of $\langle j_{ze} \rangle$ is similar to the periodic case, where the current induced by the onset of the ECDI 
at early simulation times fades away at large enough times.

On the opposite limit we have the case $u_{zi0} = 100$ km/s with $\Delta t_{i0} \ll \Delta t_{\textmd{sat}}$, where the final $E_y$ shows a sustained but very weak short-wavelength oscillation.
The azimuthal spectrum shows that dominant modes concentrate  around resonancea $m=1$ and, to a lesser extend, $m=2$. 
In this regime ECDI modes cannot develop completely since most of the ions leave the domain before full trapping occurs, so that we get an early stage of the ECDI observed with periodic conditions.
According to figure \ref{fig: jze inj}, these mild oscillatory modes induce a weak $j_{ze}$ in the electrons, which is very small compared with values produced by a fully developed ECDI.

In the intermediate regime $\Delta t_{i0} \sim \Delta t_{\textmd{sat}}$ we consider cases with $u_{zi0}$ between 5 and 25 km/s (always for hydrogen).
The plasma behavior resembles the saturated behavior of periodic simulation, because the removal and injection of particles happen at the proper rate that keeps feeding ECDI modes and allows them to fully develop, preventing the quenching of the oscillations observed in the regime $\Delta t_{i0} \gg \Delta t_{\textmd{sat}}$. 
From these cases, those with $u_{zi0} = 5$ and 25 km/s are halfway between regimes and show features of both limit and intermediate regimes.
In every case, the electric field shows sustained oscillations with a clear dominant mode close to the resonance $m=1$, that effectively induces a significant $j_{ze}$ in the long term. 
The magnitude of the induced transport depends on the case, being more significant, and comparable to the peak values of the periodic case, for the cases with $u_{zi0} = 10$ and 15 km/s.
In these two cases the dominance of modes close to $m=1$ does not happen at every time and there is some kind of intermittent or pulsed behavior of long and short wavelength modes, which results in an oscillatory evolution of the average $j_{ze}$.
This intermittency is also observed in the time evolution of figure \ref{fig: ECDI inj Ey zy} for the case $u_{zi0} = 10$ km/s.
For greater and smaller values of ion velocity,  $j_{ze}$ gets diminished, which is expected after our observations of null $j_{ze}$ in upper and lower limit regimes.

\begin{figure}[!t]
    \centering
	\begin{subfigure}[b]{0.45\textwidth}
		\caption{$u_{zi0} = 1$ km/s}
		\centering
        \includegraphics[width=\textwidth]{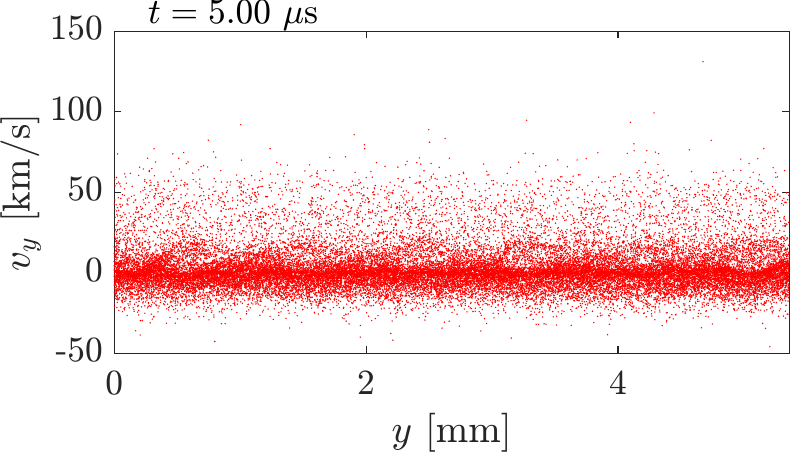}
	\end{subfigure}
    \par\bigskip
	\begin{subfigure}[b]{0.45\textwidth}
		\caption{$u_{zi0} = 10$ km/s}
		\centering
        \includegraphics[width=\textwidth]{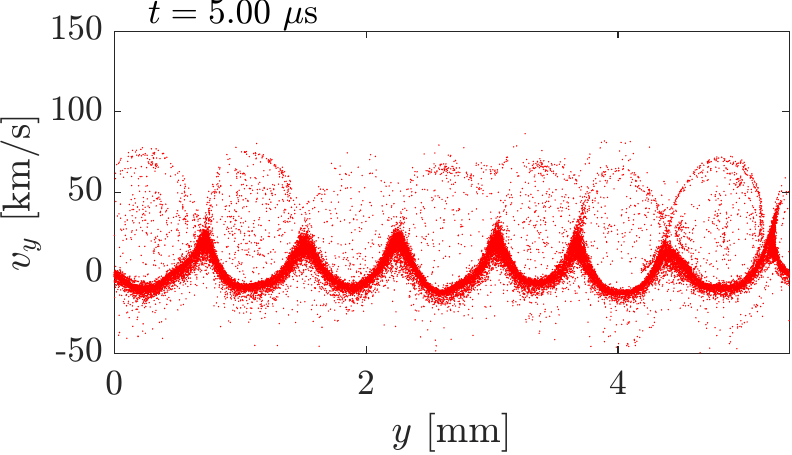}
	\end{subfigure}
    \par\bigskip
	\begin{subfigure}[b]{0.45\textwidth}
		\caption{$u_{zi0} = 25$ km/s}
		\centering
        \includegraphics[width=\textwidth]{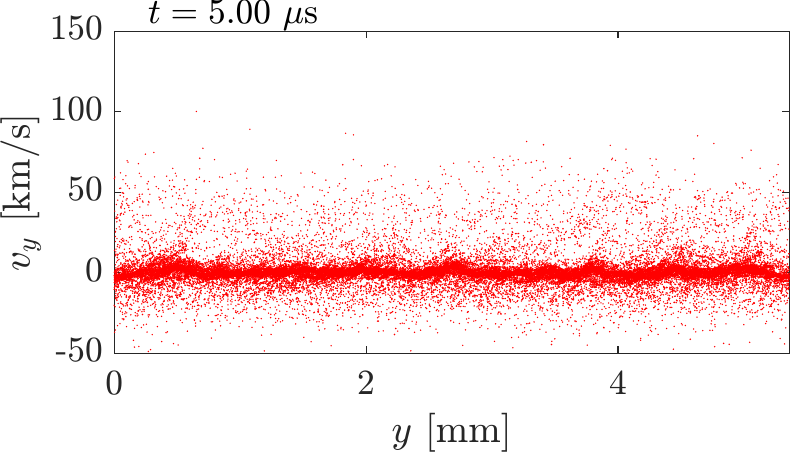}
	\end{subfigure}
    \caption{Finite plasma with axial injection.  Late behavior (at $t=5 \mu$s) in the phase space plane $(y, v_y)$ of ion particles contained in the axial slab $0.7 \leq z/L_z \leq 0.75$.
    }
    \label{fig: phase space inj}
\end{figure}

From the point of view of ion particles in phase space (see figure \ref{fig: phase space inj}), the final pictures for cases $u_{zi0} = 1$, 10 and 25 km/s are shown in an axial slab $0.7< z/L_z < 0.75$. Here, it is confirmed that only when $\Delta t_{i0} \sim \Delta t_{\textmd{sat}}$ the vortex-like structure characteristic of ion-wave trapping is preserved and long-term axial transport exists.
The existence of different regimes depending on the ion residence time could also explain why, with a virtual axial length, some groups observe a transition to an ion-acoustic mode \cite{lafl16b} 
while others do not \cite{tacc19c, smol19c}.

\begin{figure}[!t]
    \centering
	\begin{subfigure}[b]{0.49\textwidth}
		\caption{}
		\centering
        \includegraphics[width=\textwidth]{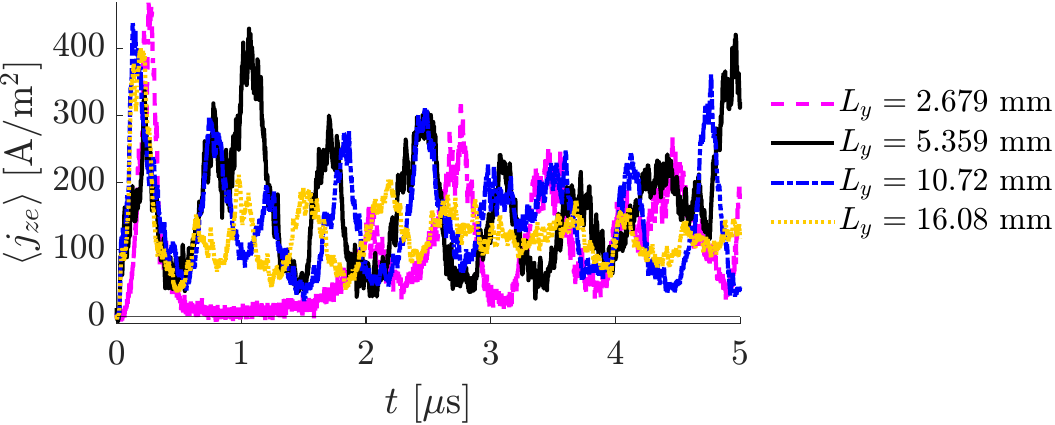}
	\end{subfigure}
	\begin{subfigure}[b]{0.49\textwidth}
		\caption{}
		\centering
        \includegraphics[width=0.8\textwidth]{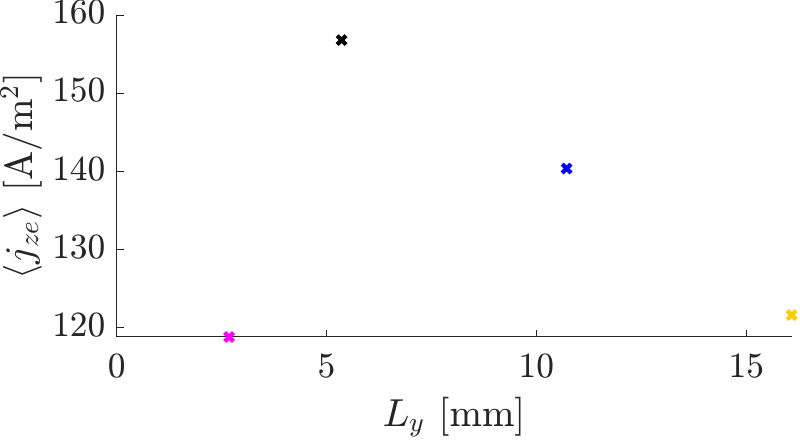}
	\end{subfigure}
     \caption{Finite plasma with axial injection. (a) Time evolution of the volume-averaged electron axial current $\angl{j_{ze}}$ for several $L_y$ and (b) its time average for $t \geq 2$ $\mu$s.
     }
    \label{fig: jze inj Ly}
\end{figure}

The sensitivity of these results to $L_y$ has been tested to ensure that the same conclusions apply to larger domains.
The evolution of electron currents are plotted in figure \ref{fig: jze inj Ly} varying $L_y$ while fixing $u_{zi0} = 10$ km/s.
As with fully periodic simulations, the shortest case gives the most different transient but in all cases a large $j_{ze}$ develops and they all show the characteristics of the intermediate regime $\Delta t_{i0} \sim \Delta t_{\textmd{sat}}$. 
In light of the fully periodic results, where larger $L_y$ seemed to favor the formation of long domain-size modes, it is surprising that, here, increasing $L_y$ mitigates the intermittency of short and long scales and yields less oscillatory currents.
A possible explanation could be that the saturation time is slightly affected by $L_y$.

\subsection{Energy balance}

For the fully periodic case, equation \eqref{eq: derivative total energy} shows that the evolution of total energy in the domain is fed by the equilibrium electric field $E_0$ and is tied to the presence of an average $j_{ze}$. Therefore, an energetically stationary state does not allow for an axial electron current. This theoretical conclusion is retrieved in our periodic simulations. 

Axial injection and removal of particles involve energy inputs and losses through axial boundaries that have to be accounted for in the energy balance, yielding
\begin{equation}
    \dertot{\mathcal E}{t} = E_0 \langle j_{ze} \rangle \mathcal V - P_{\textmd{out}},
    \label{eq: derivative total energy inj}
\end{equation}
where $P_{\textmd{out}}$ gathers the net energy outflow through axial boundaries and can be computed from the energies of removed and injected particles.
The new term opens the possibility to have a balance between the energy input by $E_0$ and  boundary losses, allowing for an energetically stationary behavior and an average $j_{ze}$ at the same time. 
Any energy loss term, such as inelastic collisions, could play a similar role in the balance.

The different terms in the balance equation \eqref{eq: derivative total energy inj} are computed and represented in figure \ref{fig: energy inj} for the case with $u_{zi0} = 10$ km/s, in the intermediate regime with a net $j_{ze}$. The balance is approximately fulfilled and the energy is close to stationary. There is, however, small changes in the total energy as well as in the partial energies of ions and electrons. The losses introduced by the axial boundary conditions limit enormously the heating that was observed in the fully periodic cases without losses. Actually, the total and species energies remain within levels close to the initial population. 
The electrostatic energy is, again, a minor contribution to the total energy.   

\section{Conclusions}

The first part of this article is focused on the simulation of the classical ECDI with the PIC formulation and settings as close as possible to the linear kinetic theory. 
Unstable short-wavelength modes are seen to grow in the initially homogeneous plasma that fit qualitatively well the features of the theoretical ECDI dispersion relation in early simulation times.
Close to saturation, some parts of the Fourier spectra show similarities with ion-acoustic modes.
After saturation, the short scale modes vanish and the plasma tends to a new equilibrium with mild or long-wavelength oscillations and much more mode mixing.
It is only during the growth and saturation of the ECDI modes that a turbulence-based axial current is induced in the electrons. 
The non-linear saturation is related with the ion distribution that yields the characteristic vortex-like structure in phase space.
The vanishing of oscillations seems related with the blurring of those vortexes.

This behavior differs from the unlimited growth reported in the literature of 1D ECDI simulations with no virtual axial length \cite{janh18,lafl16a,smol23}, where non-linear saturation or the quenching of short modes were not seen. 
It is possible that the evolution we observe is not inherent to every ECDI simulation and depends on the choice of parameters.
Other possibility is that more simulation time is needed in these works to reproduce the full behavior reported here, which is in line with our parametric analysis on the ion mass. 

When axial boundaries are replaced by removal/injection surfaces, the continuous renewal of particles can yield completely different behaviors. 
These boundary conditions imply that particles that have interacted with the electrostatic wave are removed from the simulation eventually and new particles with the original distribution are injected. 
\eb{Our approach (i.e., constant injection fluxes and 2D domain) and motivation (i.e., based on ion transport and not in non-linear saturation) are different from previous 1D works with particle refreshing \cite{tacc19c,lafl16a}}.
These simulations show that a key parameter is the ion injection velocity $u_{zi0}$, while its effect is negligible in the ECDI dispersion relation and periodic simulations. 
The intermediate regime where saturation and ion-residence times are similar is the only one yielding long-term short-scale oscillations and a turbulent-based axial current, and it is the most interesting one in the context of Hall discharges. 

Previous results can be extrapolated to xenon in a Hall discharge using the scaling laws suggested in section \ref{sec: ion mass} and a typical $\Delta t_{i0}$.
Using an average ion velocity $u_{zi0} = 17$ km/s and anode-to-cathode length $L_z = 3.35$ cm (e.g., from results in \cite{bell23}), $\Delta t_{i0}$ is estimated to be 1.95 $\mu$s.
The value of $\Delta t_{\textmd{sat}}$ for xenon mass is estimated to be 1.83 to 2.06 $\mu s$.
This similarity suggests that the intermediate regime of the ECDI in a finite plasma could develop in a conventional Hall discharge and supports the idea of the ECDI possibly being an important actor in the anomalous electron transport.
\eb{A more exhaustive analysis relating these conclusions to Hall devices is left as future work, which could start by introducing the effect of the equilibrium electric field on the ions.}

\section*{Acknowledgments}

This work has been supported by the R\&D project PID2022-140035OB-I00 (HEEP) funded by MCIN/AEI/10.13039/501100011033 and by ``ERDF A way of making Europe''.
E. Bello-Benítez thanks the financial support by the European Research Council
under the European Union’s Horizon 2020 research and
innovation programme (project ZARATHUSTRA, grant
agreement No 950466).
A. Marín-Cebrián thanks the financial support by MCIN/AEI/10.13039/501100011033 under grant FPU20/02901.
The authors acknowledge useful discussions on this work with J. J. Ramos.
This work is part of the PhD thesis of E. Bello-Benítez, under development, and it will be included as a chapter within it.


%


\begin{thebibliography}{27}%
	\makeatletter
	\providecommand \@ifxundefined [1]{%
		\@ifx{#1\undefined}
	}%
	\providecommand \@ifnum [1]{%
		\ifnum #1\expandafter \@firstoftwo
		\else \expandafter \@secondoftwo
		\fi
	}%
	\providecommand \@ifx [1]{%
		\ifx #1\expandafter \@firstoftwo
		\else \expandafter \@secondoftwo
		\fi
	}%
	\providecommand \natexlab [1]{#1}%
	\providecommand \enquote  [1]{``#1''}%
	\providecommand \bibnamefont  [1]{#1}%
	\providecommand \bibfnamefont [1]{#1}%
	\providecommand \citenamefont [1]{#1}%
	\providecommand \href@noop [0]{\@secondoftwo}%
	\providecommand \href [0]{\begingroup \@sanitize@url \@href}%
	\providecommand \@href[1]{\@@startlink{#1}\@@href}%
	\providecommand \@@href[1]{\endgroup#1\@@endlink}%
	\providecommand \@sanitize@url [0]{\catcode `\\12\catcode `\$12\catcode
		`\&12\catcode `\#12\catcode `\^12\catcode `\_12\catcode `\%12\relax}%
	\providecommand \@@startlink[1]{}%
	\providecommand \@@endlink[0]{}%
	\providecommand \url  [0]{\begingroup\@sanitize@url \@url }%
	\providecommand \@url [1]{\endgroup\@href {#1}{\urlprefix }}%
	\providecommand \urlprefix  [0]{URL }%
	\providecommand \Eprint [0]{\href }%
	\providecommand \doibase [0]{https://doi.org/}%
	\providecommand \selectlanguage [0]{\@gobble}%
	\providecommand \bibinfo  [0]{\@secondoftwo}%
	\providecommand \bibfield  [0]{\@secondoftwo}%
	\providecommand \translation [1]{[#1]}%
	\providecommand \BibitemOpen [0]{}%
	\providecommand \bibitemStop [0]{}%
	\providecommand \bibitemNoStop [0]{.\EOS\space}%
	\providecommand \EOS [0]{\spacefactor3000\relax}%
	\providecommand \BibitemShut  [1]{\csname bibitem#1\endcsname}%
	\let\auto@bib@innerbib\@empty
	\bibitem [{\citenamefont {Choueiri}\ and\ \citenamefont
		{Ziemer}(2001)}]{chou01}%
	\BibitemOpen
	\bibfield  {author} {\bibinfo {author} {\bibfnamefont {E.}~\bibnamefont
			{Choueiri}}\ and\ \bibinfo {author} {\bibfnamefont {J.}~\bibnamefont
			{Ziemer}},\ }\bibfield  {title} {\bibinfo {title} {Quasi-steady
			magnetoplasmadynamic thruster performance database},\ }\href@noop {}
	{\bibfield  {journal} {\bibinfo  {journal} {Journal of Propulsion and Power}\
		}\textbf {\bibinfo {volume} {17}},\ \bibinfo {pages} {967} (\bibinfo {year}
		{2001})}\BibitemShut {NoStop}%
	\bibitem [{\citenamefont {Ellison}\ \emph {et~al.}(2012)\citenamefont
		{Ellison}, \citenamefont {Raitses},\ and\ \citenamefont {Fisch}}]{elli12}%
	\BibitemOpen
	\bibfield  {author} {\bibinfo {author} {\bibfnamefont {C.}~\bibnamefont
			{Ellison}}, \bibinfo {author} {\bibfnamefont {Y.}~\bibnamefont {Raitses}},\
		and\ \bibinfo {author} {\bibfnamefont {N.}~\bibnamefont {Fisch}},\ }\bibfield
	{title} {\bibinfo {title} {Cross-field electron transport induced by a
			rotating spoke in a cylindrical {Hall} thruster},\ }\href@noop {} {\bibfield
		{journal} {\bibinfo  {journal} {Physics of Plasmas}\ }\textbf {\bibinfo
			{volume} {19}},\ \bibinfo {pages} {013503} (\bibinfo {year}
		{2012})}\BibitemShut {NoStop}%
	\bibitem [{\citenamefont {MacDonald}\ \emph {et~al.}(2011)\citenamefont
		{MacDonald}, \citenamefont {Cappelli}, \citenamefont {Gildea}, \citenamefont
		{Martinez-Sanchez},\ and\ \citenamefont {Hargus~Jr}}]{macd11}%
	\BibitemOpen
	\bibfield  {author} {\bibinfo {author} {\bibfnamefont {N.~A.}\ \bibnamefont
			{MacDonald}}, \bibinfo {author} {\bibfnamefont {M.~A.}\ \bibnamefont
			{Cappelli}}, \bibinfo {author} {\bibfnamefont {S.~R.}\ \bibnamefont
			{Gildea}}, \bibinfo {author} {\bibfnamefont {M.}~\bibnamefont
			{Martinez-Sanchez}},\ and\ \bibinfo {author} {\bibfnamefont {W.~A.}\
			\bibnamefont {Hargus~Jr}},\ }\bibfield  {title} {\bibinfo {title}
		{Laser-induced fluorescence velocity measurements of a diverging cusped-field
			thruster},\ }\href@noop {} {\bibfield  {journal} {\bibinfo  {journal}
			{{Journal of Physics D: Applied Physics}}\ }\textbf {\bibinfo {volume}
			{44}},\ \bibinfo {pages} {295203} (\bibinfo {year} {2011})}\BibitemShut
	{NoStop}%
	\bibitem [{\citenamefont {Tsikata}\ \emph {et~al.}(2009)\citenamefont
		{Tsikata}, \citenamefont {Lemoine}, \citenamefont {Pisarev},\ and\
		\citenamefont {Gresillon}}]{tsik09b}%
	\BibitemOpen
	\bibfield  {author} {\bibinfo {author} {\bibfnamefont {S.}~\bibnamefont
			{Tsikata}}, \bibinfo {author} {\bibfnamefont {N.}~\bibnamefont {Lemoine}},
		\bibinfo {author} {\bibfnamefont {V.}~\bibnamefont {Pisarev}},\ and\ \bibinfo
		{author} {\bibfnamefont {D.}~\bibnamefont {Gresillon}},\ }\bibfield  {title}
	{\bibinfo {title} {Dispersion relations of electron density fluctuations in a
			{H}all thruster plasma, observed by collective light scattering},\
	}\href@noop {} {\bibfield  {journal} {\bibinfo  {journal} {Physics of
				Plasmas}\ }\textbf {\bibinfo {volume} {16}},\ \bibinfo {pages} {033506}
		(\bibinfo {year} {2009})}\BibitemShut {NoStop}%
	\bibitem [{\citenamefont {Tsikata}\ and\ \citenamefont {Minea}(2015)}]{tsik15}%
	\BibitemOpen
	\bibfield  {author} {\bibinfo {author} {\bibfnamefont {S.}~\bibnamefont
			{Tsikata}}\ and\ \bibinfo {author} {\bibfnamefont {T.}~\bibnamefont
			{Minea}},\ }\bibfield  {title} {\bibinfo {title} {Modulated electron
			cyclotron drift instability in a high-power pulsed magnetron discharge},\
	}\href@noop {} {\bibfield  {journal} {\bibinfo  {journal} {Physical Review
				Letters}\ }\textbf {\bibinfo {volume} {114}},\ \bibinfo {pages} {185001}
		(\bibinfo {year} {2015})}\BibitemShut {NoStop}%
	\bibitem [{\citenamefont {Tsikata}\ \emph {et~al.}(2017)\citenamefont
		{Tsikata}, \citenamefont {H\'eron},\ and\ \citenamefont
		{Honor\'e}}]{tsik19c}%
	\BibitemOpen
	\bibfield  {author} {\bibinfo {author} {\bibfnamefont {S.}~\bibnamefont
			{Tsikata}}, \bibinfo {author} {\bibfnamefont {A.}~\bibnamefont {H\'eron}},\
		and\ \bibinfo {author} {\bibfnamefont {C.}~\bibnamefont {Honor\'e}},\
	}\bibfield  {title} {\bibinfo {title} {Hall thruster microturbulence under
			conditions of modified electron wall emission},\ }\href
	{https://doi.org/10.1063/1.4984255} {\bibfield  {journal} {\bibinfo
			{journal} {Physics of Plasmas}\ }\textbf {\bibinfo {volume} {24}},\ \bibinfo
		{pages} {053519} (\bibinfo {year} {2017})}\BibitemShut {NoStop}%
	\bibitem [{\citenamefont {Janes}\ and\ \citenamefont {Lowder}(1966)}]{jane66}%
	\BibitemOpen
	\bibfield  {author} {\bibinfo {author} {\bibfnamefont {G.}~\bibnamefont
			{Janes}}\ and\ \bibinfo {author} {\bibfnamefont {R.}~\bibnamefont {Lowder}},\
	}\bibfield  {title} {\bibinfo {title} {Anomalous electron diffusion and ion
			acceleration in a low-density plasma},\ }\href@noop {} {\bibfield  {journal}
		{\bibinfo  {journal} {Physics of Fluids}\ }\textbf {\bibinfo {volume} {9}},\
		\bibinfo {pages} {1115} (\bibinfo {year} {1966})}\BibitemShut {NoStop}%
	\bibitem [{\citenamefont {Forslund}\ \emph {et~al.}(1970)\citenamefont
		{Forslund}, \citenamefont {Morse},\ and\ \citenamefont {Nielson}}]{fors70}%
	\BibitemOpen
	\bibfield  {author} {\bibinfo {author} {\bibfnamefont {D.}~\bibnamefont
			{Forslund}}, \bibinfo {author} {\bibfnamefont {R.}~\bibnamefont {Morse}},\
		and\ \bibinfo {author} {\bibfnamefont {C.}~\bibnamefont {Nielson}},\
	}\bibfield  {title} {\bibinfo {title} {Electron cyclotron drift
			instability},\ }\href@noop {} {\bibfield  {journal} {\bibinfo  {journal}
			{Physical Review Letters}\ }\textbf {\bibinfo {volume} {25}},\ \bibinfo
		{pages} {1266} (\bibinfo {year} {1970})}\BibitemShut {NoStop}%
	\bibitem [{\citenamefont {Wong}(1970)}]{wong70}%
	\BibitemOpen
	\bibfield  {author} {\bibinfo {author} {\bibfnamefont {H.}~\bibnamefont
			{Wong}},\ }\bibfield  {title} {\bibinfo {title} {Electrostatic electron-ion
			streaming instability},\ }\href@noop {} {\bibfield  {journal} {\bibinfo
			{journal} {Physics of Fluids}\ }\textbf {\bibinfo {volume} {13}},\ \bibinfo
		{pages} {757} (\bibinfo {year} {1970})}\BibitemShut {NoStop}%
	\bibitem [{\citenamefont {Adam}\ \emph {et~al.}(2004)\citenamefont {Adam},
		\citenamefont {Her{\'o}n},\ and\ \citenamefont {Laval}}]{adam04}%
	\BibitemOpen
	\bibfield  {author} {\bibinfo {author} {\bibfnamefont {J.}~\bibnamefont
			{Adam}}, \bibinfo {author} {\bibfnamefont {A.}~\bibnamefont {Her{\'o}n}},\
		and\ \bibinfo {author} {\bibfnamefont {G.}~\bibnamefont {Laval}},\ }\bibfield
	{title} {\bibinfo {title} {Study of stationary plasma thrusters using
			two-dimensional fully kinetic simulations},\ }\href@noop {} {\bibfield
		{journal} {\bibinfo  {journal} {Physics of Plasmas}\ }\textbf {\bibinfo
			{volume} {11}},\ \bibinfo {pages} {295} (\bibinfo {year} {2004})}\BibitemShut
	{NoStop}%
	\bibitem [{\citenamefont {Ducrocq}\ \emph {et~al.}(2006)\citenamefont
		{Ducrocq}, \citenamefont {Adam}, \citenamefont {H{\'e}ron},\ and\
		\citenamefont {Laval}}]{ducr06}%
	\BibitemOpen
	\bibfield  {author} {\bibinfo {author} {\bibfnamefont {A.}~\bibnamefont
			{Ducrocq}}, \bibinfo {author} {\bibfnamefont {J.}~\bibnamefont {Adam}},
		\bibinfo {author} {\bibfnamefont {A.}~\bibnamefont {H{\'e}ron}},\ and\
		\bibinfo {author} {\bibfnamefont {G.}~\bibnamefont {Laval}},\ }\bibfield
	{title} {\bibinfo {title} {High-frequency electron drift instability in the
			cross-field configuration of {Hall} thrusters},\ }\href@noop {} {\bibfield
		{journal} {\bibinfo  {journal} {Physics of Plasmas}\ }\textbf {\bibinfo
			{volume} {13}},\ \bibinfo {pages} {102111} (\bibinfo {year}
		{2006})}\BibitemShut {NoStop}%
	\bibitem [{\citenamefont {Cavalier}\ \emph {et~al.}(2013)\citenamefont
		{Cavalier}, \citenamefont {Lemoine}, \citenamefont {Bonhomme}, \citenamefont
		{Tsikata}, \citenamefont {Honoré},\ and\ \citenamefont
		{Grésillon}}]{cava13}%
	\BibitemOpen
	\bibfield  {author} {\bibinfo {author} {\bibfnamefont {J.}~\bibnamefont
			{Cavalier}}, \bibinfo {author} {\bibfnamefont {N.}~\bibnamefont {Lemoine}},
		\bibinfo {author} {\bibfnamefont {G.}~\bibnamefont {Bonhomme}}, \bibinfo
		{author} {\bibfnamefont {S.}~\bibnamefont {Tsikata}}, \bibinfo {author}
		{\bibfnamefont {C.}~\bibnamefont {Honoré}},\ and\ \bibinfo {author}
		{\bibfnamefont {D.}~\bibnamefont {Grésillon}},\ }\bibfield  {title}
	{\bibinfo {title} {Hall thruster plasma fluctuations identified as the exb
			electron drift instability: Modeling and fitting on experimental data},\
	}\href@noop {} {\bibfield  {journal} {\bibinfo  {journal} {PoP}\ }\textbf
		{\bibinfo {volume} {20}},\ \bibinfo {pages} {082107} (\bibinfo {year}
		{2013})}\BibitemShut {NoStop}%
	\bibitem [{\citenamefont {Lafleur}\ \emph
		{et~al.}(2016{\natexlab{a}})\citenamefont {Lafleur}, \citenamefont
		{Baalrud},\ and\ \citenamefont {Chabert}}]{lafl16b}%
	\BibitemOpen
	\bibfield  {author} {\bibinfo {author} {\bibfnamefont {T.}~\bibnamefont
			{Lafleur}}, \bibinfo {author} {\bibfnamefont {S.}~\bibnamefont {Baalrud}},\
		and\ \bibinfo {author} {\bibfnamefont {P.}~\bibnamefont {Chabert}},\
	}\bibfield  {title} {\bibinfo {title} {Theory for the anomalous electron
			transport in {H}all effect thrusters. ii. kinetic model},\ }\href@noop {}
	{\bibfield  {journal} {\bibinfo  {journal} {Physics of Plasmas}\ }\textbf
		{\bibinfo {volume} {23}},\ \bibinfo {pages} {053503} (\bibinfo {year}
		{2016}{\natexlab{a}})}\BibitemShut {NoStop}%
	\bibitem [{\citenamefont {Janhunen}\ \emph
		{et~al.}(2018{\natexlab{a}})\citenamefont {Janhunen}, \citenamefont
		{Smolyakov}, \citenamefont {Sydorenko}, \citenamefont {Jimenez},
		\citenamefont {Kaganovich},\ and\ \citenamefont {Raitses}}]{janh18b}%
	\BibitemOpen
	\bibfield  {author} {\bibinfo {author} {\bibfnamefont {S.}~\bibnamefont
			{Janhunen}}, \bibinfo {author} {\bibfnamefont {A.}~\bibnamefont {Smolyakov}},
		\bibinfo {author} {\bibfnamefont {D.}~\bibnamefont {Sydorenko}}, \bibinfo
		{author} {\bibfnamefont {M.}~\bibnamefont {Jimenez}}, \bibinfo {author}
		{\bibfnamefont {I.}~\bibnamefont {Kaganovich}},\ and\ \bibinfo {author}
		{\bibfnamefont {Y.}~\bibnamefont {Raitses}},\ }\bibfield  {title} {\bibinfo
		{title} {Evolution of the electron cyclotron drift instability in
			two-dimensions},\ }\href@noop {} {\bibfield  {journal} {\bibinfo  {journal}
			{Physics of Plasmas}\ }\textbf {\bibinfo {volume} {25}},\ \bibinfo {pages}
		{082308} (\bibinfo {year} {2018}{\natexlab{a}})}\BibitemShut {NoStop}%
	\bibitem [{\citenamefont {Janhunen}\ \emph
		{et~al.}(2018{\natexlab{b}})\citenamefont {Janhunen}, \citenamefont
		{Smolyakov}, \citenamefont {Chapurin}, \citenamefont {Sydorenko},
		\citenamefont {Kaganovich},\ and\ \citenamefont {Raitses}}]{janh18}%
	\BibitemOpen
	\bibfield  {author} {\bibinfo {author} {\bibfnamefont {S.}~\bibnamefont
			{Janhunen}}, \bibinfo {author} {\bibfnamefont {A.}~\bibnamefont {Smolyakov}},
		\bibinfo {author} {\bibfnamefont {O.}~\bibnamefont {Chapurin}}, \bibinfo
		{author} {\bibfnamefont {D.}~\bibnamefont {Sydorenko}}, \bibinfo {author}
		{\bibfnamefont {I.}~\bibnamefont {Kaganovich}},\ and\ \bibinfo {author}
		{\bibfnamefont {Y.}~\bibnamefont {Raitses}},\ }\bibfield  {title} {\bibinfo
		{title} {Nonlinear structures and anomalous transport in partially magnetized
			{ExB} plasmas},\ }\href@noop {} {\bibfield  {journal} {\bibinfo  {journal}
			{Physics of Plasmas}\ }\textbf {\bibinfo {volume} {25}},\ \bibinfo {pages}
		{11608} (\bibinfo {year} {2018}{\natexlab{b}})}\BibitemShut {NoStop}%
	\bibitem [{\citenamefont {Lafleur}\ \emph
		{et~al.}(2016{\natexlab{b}})\citenamefont {Lafleur}, \citenamefont
		{Baalrud},\ and\ \citenamefont {Chabert}}]{lafl16a}%
	\BibitemOpen
	\bibfield  {author} {\bibinfo {author} {\bibfnamefont {T.}~\bibnamefont
			{Lafleur}}, \bibinfo {author} {\bibfnamefont {S.}~\bibnamefont {Baalrud}},\
		and\ \bibinfo {author} {\bibfnamefont {P.}~\bibnamefont {Chabert}},\
	}\bibfield  {title} {\bibinfo {title} {Theory for the anomalous electron
			transport in hall effect thrusters. i. insights from particle-in-cell
			simulations},\ }\href@noop {} {\bibfield  {journal} {\bibinfo  {journal}
			{Physics of Plasmas}\ }\textbf {\bibinfo {volume} {23}},\ \bibinfo {pages}
		{053502} (\bibinfo {year} {2016}{\natexlab{b}})}\BibitemShut {NoStop}%
	\bibitem [{\citenamefont {Asadi}\ \emph {et~al.}(2019)\citenamefont {Asadi},
		\citenamefont {Taccogna},\ and\ \citenamefont {Sharifian}}]{tacc19c}%
	\BibitemOpen
	\bibfield  {author} {\bibinfo {author} {\bibfnamefont {Z.}~\bibnamefont
			{Asadi}}, \bibinfo {author} {\bibfnamefont {F.}~\bibnamefont {Taccogna}},\
		and\ \bibinfo {author} {\bibfnamefont {M.}~\bibnamefont {Sharifian}},\
	}\bibfield  {title} {\bibinfo {title} {Numerical study of electron cyclotron
			drift instability: Application to hall thruster},\ }\bibfield  {journal}
	{\bibinfo  {journal} {Frontiers in Physics}\ }\textbf {\bibinfo {volume}
		{7}},\ \href {https://doi.org/10.3389/fphy.2019.00140}
	{10.3389/fphy.2019.00140} (\bibinfo {year} {2019})\BibitemShut {NoStop}%
	\bibitem [{\citenamefont {Lafleur}\ and\ \citenamefont
		{Chabert}(2017)}]{lafl17b}%
	\BibitemOpen
	\bibfield  {author} {\bibinfo {author} {\bibfnamefont {T.}~\bibnamefont
			{Lafleur}}\ and\ \bibinfo {author} {\bibfnamefont {P.}~\bibnamefont
			{Chabert}},\ }\bibfield  {title} {\bibinfo {title} {{The role of
				instability-enhanced friction on `anomalous' electron and ion transport in
				{H}all-effect thrusters}},\ }\href@noop {} {\bibfield  {journal} {\bibinfo
			{journal} {Plasma Sources Science and Technology}\ }\textbf {\bibinfo
			{volume} {27}},\ \bibinfo {pages} {015003} (\bibinfo {year}
		{2017})}\BibitemShut {NoStop}%
	\bibitem [{\citenamefont {Coche}\ and\ \citenamefont
		{Garrigues}(2014)}]{coche14}%
	\BibitemOpen
	\bibfield  {author} {\bibinfo {author} {\bibfnamefont {P.}~\bibnamefont
			{Coche}}\ and\ \bibinfo {author} {\bibfnamefont {L.}~\bibnamefont
			{Garrigues}},\ }\bibfield  {title} {\bibinfo {title} {A two-dimensional
			(azimuthal-axial) particle-in-cell model of a hall thruster},\ }\href@noop {}
	{\bibfield  {journal} {\bibinfo  {journal} {Physics of Plasmas}\ }\textbf
		{\bibinfo {volume} {21}},\ \bibinfo {pages} {023503} (\bibinfo {year}
		{2014})}\BibitemShut {NoStop}%
	\bibitem [{\citenamefont {Smolyakov}\ \emph {et~al.}(2019)\citenamefont
		{Smolyakov}, \citenamefont {Zintel}, \citenamefont {Couedel}, \citenamefont
		{Sydorenko}, \citenamefont {Umnov}, \citenamefont {Sorokina},\ and\
		\citenamefont {Marusov}}]{smol19c}%
	\BibitemOpen
	\bibfield  {author} {\bibinfo {author} {\bibfnamefont {A.}~\bibnamefont
			{Smolyakov}}, \bibinfo {author} {\bibfnamefont {T.}~\bibnamefont {Zintel}},
		\bibinfo {author} {\bibfnamefont {L.}~\bibnamefont {Couedel}}, \bibinfo
		{author} {\bibfnamefont {D.}~\bibnamefont {Sydorenko}}, \bibinfo {author}
		{\bibfnamefont {A.}~\bibnamefont {Umnov}}, \bibinfo {author} {\bibfnamefont
			{E.}~\bibnamefont {Sorokina}},\ and\ \bibinfo {author} {\bibfnamefont
			{N.}~\bibnamefont {Marusov}},\ }\bibfield  {title} {\bibinfo {title}
		{Anomalous electron transport in one-dimensional electron cyclotron drift
			turbulence},\ }\href@noop {} {\bibfield  {journal} {\bibinfo  {journal}
			{Plasma Physics Reports}\ }\textbf {\bibinfo {volume} {46}} (\bibinfo {year}
		{2019})}\BibitemShut {NoStop}%
	\bibitem [{\citenamefont {Tavassoli}\ \emph {et~al.}(2023)\citenamefont
		{Tavassoli}, \citenamefont {Papahn~Zadeh}, \citenamefont {Smolyakov},
		\citenamefont {Shoucri},\ and\ \citenamefont {Spiteri}}]{smol23}%
	\BibitemOpen
	\bibfield  {author} {\bibinfo {author} {\bibfnamefont {A.}~\bibnamefont
			{Tavassoli}}, \bibinfo {author} {\bibfnamefont {M.}~\bibnamefont
			{Papahn~Zadeh}}, \bibinfo {author} {\bibfnamefont {A.}~\bibnamefont
			{Smolyakov}}, \bibinfo {author} {\bibfnamefont {M.}~\bibnamefont {Shoucri}},\
		and\ \bibinfo {author} {\bibfnamefont {R.~J.}\ \bibnamefont {Spiteri}},\
	}\bibfield  {title} {\bibinfo {title} {The electron cyclotron drift
			instability: A comparison of particle-in-cell and continuum {V}lasov
			simulations},\ }\bibfield  {journal} {\bibinfo  {journal} {Physics of
			Plasmas}\ }\textbf {\bibinfo {volume} {30}},\ \href
	{https://doi.org/10.1063/5.0134457} {10.1063/5.0134457} (\bibinfo {year}
	{2023}),\ \bibinfo {note} {033905}\BibitemShut {NoStop}%
	\bibitem [{\citenamefont {Swanson}(2003)}]{swan03}%
	\BibitemOpen
	\bibfield  {author} {\bibinfo {author} {\bibfnamefont {D.}~\bibnamefont
			{Swanson}},\ }\href@noop {} {\emph {\bibinfo {title} {Plasma {W}aves, 2nd
				{E}dition}}}\ (\bibinfo  {publisher} {IOP Publishing, Bristol, UK},\ \bibinfo
	{year} {2003})\BibitemShut {NoStop}%
	\bibitem [{\citenamefont {Frigo}\ and\ \citenamefont {Johnson}(2005)}]{frig05}%
	\BibitemOpen
	\bibfield  {author} {\bibinfo {author} {\bibfnamefont {M.}~\bibnamefont
			{Frigo}}\ and\ \bibinfo {author} {\bibfnamefont {S.~G.}\ \bibnamefont
			{Johnson}},\ }\bibfield  {title} {\bibinfo {title} {The design and
			implementation of {FFTW3}},\ }\href@noop {} {\bibfield  {journal} {\bibinfo
			{journal} {Proceedings of the IEEE}\ }\textbf {\bibinfo {volume} {93}},\
		\bibinfo {pages} {216} (\bibinfo {year} {2005})}\BibitemShut {NoStop}%
	\bibitem [{\citenamefont {Tavassoli}\ \emph {et~al.}(2021)\citenamefont
		{Tavassoli}, \citenamefont {Chapurin}, \citenamefont {Jimenez}, \citenamefont
		{Papahn~Zadeh}, \citenamefont {Zintel}, \citenamefont {Sengupta},
		\citenamefont {Couëdel}, \citenamefont {Spiteri}, \citenamefont {Shoucri},\
		and\ \citenamefont {Smolyakov}}]{smol21}%
	\BibitemOpen
	\bibfield  {author} {\bibinfo {author} {\bibfnamefont {A.}~\bibnamefont
			{Tavassoli}}, \bibinfo {author} {\bibfnamefont {O.}~\bibnamefont {Chapurin}},
		\bibinfo {author} {\bibfnamefont {M.}~\bibnamefont {Jimenez}}, \bibinfo
		{author} {\bibfnamefont {M.}~\bibnamefont {Papahn~Zadeh}}, \bibinfo {author}
		{\bibfnamefont {T.}~\bibnamefont {Zintel}}, \bibinfo {author} {\bibfnamefont
			{M.}~\bibnamefont {Sengupta}}, \bibinfo {author} {\bibfnamefont
			{L.}~\bibnamefont {Couëdel}}, \bibinfo {author} {\bibfnamefont {R.~J.}\
			\bibnamefont {Spiteri}}, \bibinfo {author} {\bibfnamefont {M.}~\bibnamefont
			{Shoucri}},\ and\ \bibinfo {author} {\bibfnamefont {A.}~\bibnamefont
			{Smolyakov}},\ }\bibfield  {title} {\bibinfo {title} {The role of noise in
			pic and {V}lasov simulations of the {B}uneman instability},\ }\bibfield
	{journal} {\bibinfo  {journal} {Physics of Plasmas}\ }\textbf {\bibinfo
		{volume} {28}},\ \href {https://doi.org/10.1063/5.0070482}
	{10.1063/5.0070482} (\bibinfo {year} {2021})\BibitemShut {NoStop}%
	\bibitem [{\citenamefont {Charoy}\ \emph {et~al.}(2019)\citenamefont {Charoy},
		\citenamefont {Boeuf}, \citenamefont {Bourdon}, \citenamefont {Carlsson},
		\citenamefont {Chabert}, \citenamefont {Cuenot}, \citenamefont {Eremin},
		\citenamefont {Garrigues}, \citenamefont {Hara}, \citenamefont {Kaganovich},
		\citenamefont {Powis}, \citenamefont {Smolyakov}, \citenamefont {Sydorenko},
		\citenamefont {Tavant}, \citenamefont {Vermorel},\ and\ \citenamefont
		{Villafana}}]{char19b}%
	\BibitemOpen
	\bibfield  {author} {\bibinfo {author} {\bibfnamefont {T.}~\bibnamefont
			{Charoy}}, \bibinfo {author} {\bibfnamefont {J.~P.}\ \bibnamefont {Boeuf}},
		\bibinfo {author} {\bibfnamefont {A.}~\bibnamefont {Bourdon}}, \bibinfo
		{author} {\bibfnamefont {J.~A.}\ \bibnamefont {Carlsson}}, \bibinfo {author}
		{\bibfnamefont {P.}~\bibnamefont {Chabert}}, \bibinfo {author} {\bibfnamefont
			{B.}~\bibnamefont {Cuenot}}, \bibinfo {author} {\bibfnamefont
			{D.}~\bibnamefont {Eremin}}, \bibinfo {author} {\bibfnamefont
			{L.}~\bibnamefont {Garrigues}}, \bibinfo {author} {\bibfnamefont
			{K.}~\bibnamefont {Hara}}, \bibinfo {author} {\bibfnamefont {I.~D.}\
			\bibnamefont {Kaganovich}}, \bibinfo {author} {\bibfnamefont {A.~T.}\
			\bibnamefont {Powis}}, \bibinfo {author} {\bibfnamefont {A.}~\bibnamefont
			{Smolyakov}}, \bibinfo {author} {\bibfnamefont {D.}~\bibnamefont
			{Sydorenko}}, \bibinfo {author} {\bibfnamefont {A.}~\bibnamefont {Tavant}},
		\bibinfo {author} {\bibfnamefont {O.}~\bibnamefont {Vermorel}},\ and\
		\bibinfo {author} {\bibfnamefont {W.}~\bibnamefont {Villafana}},\ }\bibfield
	{title} {\bibinfo {title} {2d axial-azimuthal particle-in-cell benchmark for
			low-temperature partially magnetized plasmas},\ }\href@noop {} {\bibfield
		{journal} {\bibinfo  {journal} {Plasma Sources Science and Technology}\
		}\textbf {\bibinfo {volume} {28}},\ \bibinfo {pages} {105010} (\bibinfo
		{year} {2019})}\BibitemShut {NoStop}%
	\bibitem [{\citenamefont {Chen}\ \emph {et~al.}(2024)\citenamefont {Chen},
		\citenamefont {Kan}, \citenamefont {Gao}, \citenamefont {Duan}, \citenamefont
		{Chen}, \citenamefont {Tan},\ and\ \citenamefont {Cui}}]{chen23}%
	\BibitemOpen
	\bibfield  {author} {\bibinfo {author} {\bibfnamefont {L.}~\bibnamefont
			{Chen}}, \bibinfo {author} {\bibfnamefont {Z.-C.}\ \bibnamefont {Kan}},
		\bibinfo {author} {\bibfnamefont {W.-F.}\ \bibnamefont {Gao}}, \bibinfo
		{author} {\bibfnamefont {P.}~\bibnamefont {Duan}}, \bibinfo {author}
		{\bibfnamefont {J.-Y.}\ \bibnamefont {Chen}}, \bibinfo {author}
		{\bibfnamefont {C.-Q.}\ \bibnamefont {Tan}},\ and\ \bibinfo {author}
		{\bibfnamefont {Z.-J.}\ \bibnamefont {Cui}},\ }\bibfield  {title} {\bibinfo
		{title} {Growth mechanism and characteristics of electron drift instability
			in {Hall} thruster with different propellant types},\ }\href
	{https://doi.org/10.1088/1674-1056/acf9e5} {\bibfield  {journal} {\bibinfo
			{journal} {Chinese Physics B}\ }\textbf {\bibinfo {volume} {33}},\ \bibinfo
		{pages} {015203} (\bibinfo {year} {2024})}\BibitemShut {NoStop}%
	\bibitem [{\citenamefont {Bello-Benítez}\ and\ \citenamefont
		{Ahedo}(2023)}]{bell23}%
	\BibitemOpen
	\bibfield  {author} {\bibinfo {author} {\bibfnamefont {E.}~\bibnamefont
			{Bello-Benítez}}\ and\ \bibinfo {author} {\bibfnamefont {E.}~\bibnamefont
			{Ahedo}},\ }\bibfield  {title} {\bibinfo {title} {Stationary axial model of
			the {H}all thruster plasma discharge: electron azimuthal inertia and far
			plume effects},\ }\href {https://doi.org/10.1088/1361-6595/ad066f} {\bibfield
		{journal} {\bibinfo  {journal} {Plasma Sources Science and Technology}\
		}\textbf {\bibinfo {volume} {32}},\ \bibinfo {pages} {115011} (\bibinfo
		{year} {2023})}\BibitemShut {NoStop}%
\end{thebibliography}
\end{document}